\documentclass[12pt, draftclsnofoot, onecolumn]{IEEEtran}

\usepackage{textcomp}
\usepackage{cite}
\usepackage{graphicx,amsmath,amssymb,tikz,psfrag}
\usepackage{color}
\usepackage{fancyhdr}
\usepackage[noend]{algpseudocode}
\usepackage{parskip}
\usepackage{amsthm}
\usepackage[ruled,vlined]{algorithm2e}

\def\BibTeX{{\rm B\kern-.05em{\sc i\kern-.025em b}\kern-.08em
    T\kern-.1667em\lower.7ex\hbox{E}\kern-.125emX}}

\begin{document}

\title{Achievable DoF Regions of Three-User MIMO Broadcast Channel with  Delayed CSIT
}

\author{Tong~Zhang, and Rui Wang
\thanks{T. Zhang is with the Department of Electrical and Electronic Engineering, Southern University of Science and Technology, Shenzhen, China (email: bennyzhangtong@yahoo.com). The work of T. Zhang was done at The Chinese University of Hong Kong.

R. Wang is with the Department of Electrical and Electronic Engineering, Southern University of Science and Technology, Shenzhen, China (email:wang.r@sustech.edu.cn).}
}

\maketitle

\begin{abstract}
For the two-user multiple-input multiple-output (MIMO) broadcast channel with delayed channel state information at the transmitter (CSIT) and arbitrary antenna configurations, all the degrees-of-freedom (DoF) regions are obtained. However, for the three-user MIMO broadcast channel with delayed CSIT and arbitrary antenna configurations, the DoF region of order-2 messages is still unclear and only a partial achievable DoF region of order-1 messages is obtained, where the order-2 messages and order-1 messages are desired by two receivers and one receiver, respectively.  In this paper, for the three-user MIMO broadcast channel with delayed CSIT and arbitrary antenna configurations, we first design transmission schemes for order-2 messages and order-1 messages. Next, we propose to analyze the achievable DoF region of transmission scheme by transformation approach. In particular, we transform the decoding condition of transmission scheme w.r.t. phase duration into the achievable DoF region w.r.t. achievable DoF, through achievable DoF tuple expression connecting phase duration and achievable DoF. As a result, the DoF region of order-2 messages is characterized and an achievable DoF region of order-1 messages is completely expressed. Besides, for order-1 messages, we derive the sufficient condition, under which the proposed achievable DoF region is the DoF region.
\end{abstract}


\section{Introduction}

The multiple-input multiple-output (MIMO) broadcast channel has one transmitter and multiple receivers, all of them are equipped with multiple antennas. For the MIMO broadcast channel, the transmitter sends private messages or common messages to  receivers, where the message desired by $j=1,2,\cdots,K$ receivers are denoted by the order-$j$ message. As a fundamental metric, the degrees-of-freedom (DoF) denotes the maximal number of interference-free channels that a communication system has in high signal-to-noise ratio (SNR) regime, which is a first-order approximation of channel capacity in high SNR regime. Usually, the DoF is derived based on the match of DoF converse and achievable DoF. The DoF converse is an information-theoretic upper/outer limit, while the achievable DoF is an attainable lower/inner limit. The achievable DoF represents the number of achievable interference-free channels, which is a first-order approximation of data rate in high SNR regime. Moreover, the value of achievable DoF is also embodied in the corresponding transmission schemes for practical use.  The DoF region of the MIMO broadcast channel was obtained in \cite{1,2,3} when the channel state information at the transmitter (CSIT) is instantaneously obtained. However, when the wireless channel is fast time-varying, the instantaneous CSIT requires a high feedback frequency and a rapid feedback link, which are difficult to satisfy. To resolve that difficulty, utilizing delayed CSIT can alleviate the stringent requirements of feedback frequency and rapidity \cite{4,5,6,7,8,9,11,12,13,14,15,16,17,18,19,20,25,26,27,28}.

\subsection{Related Work}

The research on DoF of MIMO broadcast channel with delayed CSIT can be found in \cite{4,5,6,7,8,9}. As a seminal work, in \cite{4}, Maddah-Ali and Tse first characterized the DoF region of order-$j=1,2,\cdots,K$ messages for $M \ge K-j+1$ antenna configurations in the $K$-user multiple-input and single-output (MISO), i.e., the transmitter has $M$ antennas and each receiver has single antenna. Thereafter, for the two-user MIMO broadcast channel with arbitrary antenna configurations, the DoF region was derived in \cite{5,6}. For the three-user MIMO broadcast channel with symmetric antenna configurations, i.e., the transmitter has $M$ antennas and each receiver has $N$ antennas, the DoF characterization was investigated in \cite{7,8}. In \cite{7}, the sum-DoF of order-$1$ messages was obtained for $M \le N$ and  $2N \le M$ antenna configurations, and an achievable sum-DoF was derived for $N < M < 2N$ antenna configurations. In \cite{8}, for $N < M < 2.5N$ antenna configurations, we proposed a higher achievable sum-DoF than that in \cite{7} by a holistic higher-order symbol generation. For the three-user MIMO broadcast channel with arbitrary antenna configurations, an achievable DoF region of order-1 messages for $M \le \max\{N_1 + N_2, N_3\}$ antenna configurations was derived in \cite{9}. Furthermore, the study of \cite{9} showed that the achievable DoF region is the DoF region if a sufficient condition holds.

Aside from the DoF of MIMO broadcast channel with delayed CSIT, there are several related research trends \cite{11,12,13,14,15,16,17,18,19,20,25,26,27,28}. One trend is to investigate the DoF of MIMO broadcast channel with interplay of current, delayed, and no CSIT \cite{11,12,13,14,15,16,17,18,19,20}. In particular, when each receiver has either current, delayed, or no CSIT, i.e., hybrid CSIT, the DoF of MIMO broadcast channel was studied in \cite{11,12,13}. For alternating current, delayed, and no CSIT in time for each receiver, i.e., alternating CSIT, the DoF  of MIMO broadcast channel was investigated in \cite{14,15}. Under delayed and current CSIT, i.e., moderately delayed CSIT, a space-time interference alignment scheme for $K$-user MISO broadcast channel  was proposed in \cite{16}. Under delayed and imperfect current CSIT, i.e., mixed CSIT, the DoF region of two-user MISO broadcast channel  was derived in \cite{17,18}. Moreover, the DoF region of MIMO interference channel with mixed CSIT was characterized in \cite{19}. A recent study in \cite{20} investigated the sum-DoF of $K$-user MISO broadcast channel with mixed CSIT. On the other hand, the optimization of data rate for the MISO broadcast channel with delayed CSIT was investigated in \cite{25,26,27}. For the cached-aided MIMO broadcast channel with delayed CSIT, the memory-DoF tradeoff was studied in \cite{28}.

However, the DoF region of three-user MIMO broadcast channel with delayed CSIT is still an open research problem, which is the focus of this paper.

\subsection{Contributions}

In this paper, we investigate the DoF region of three-user MIMO broadcast channel with delayed CSIT and arbitrary antenna configurations. The contributions are summarized as follows:

\begin{itemize}	
 
    \item \textbf{DoF Region of Order-2 Messages}: For order-2 messages, we characterize the DoF region. For converse, we first present the DoF region with no CSIT, and then derive the DoF outer region with delayed CSIT. For   achievability, we first design the transmission scheme, and then obtain the achievable DoF region of transmission scheme by transformation approach.   
    
 	\item \textbf{Achievable DoF Region of Order-1 Messages}: For order-1 messages, since the achievable DoF region was derived for $M \le \max\{N_1 + N_2, N_3\}$ only, we obtain an achievable DoF region for $\max\{N_1 + N_2, N_3\} < M$, and a sufficient condition, under which the achievable DoF region is the DoF region. To be specific, for $\max\{N_1 + N_2, N_3\} < M \le N_2+N_3$, we design transmission schemes, where the higher-order symbol generation has a sequential manner as that in \cite{7,9}. For $N_2+N_3 < M$, we design a transmission scheme, which generalizes our holistic design of higher-order symbol generation in \cite{8} to that with arbitrary antenna configurations. Based on proposed transmission schemes, we derive the achievable DoF region and the sufficient condition of optimality by transformation approach. 
 	 	
	\item \textbf{Transformation Approach}: For the three-user MIMO broadcast channel with delayed CSIT, we propose to analyze the achievable DoF region of transmission scheme by transformation, which overcomes the drawback of achievable DoF region analysis approach in \cite{9}. As for \cite{9}, the decoding condition of transmission scheme was used to check whether the DoF outer region is achieved or not. Hence, the achievable DoF region is hard to derive, when the DoF outer region is not attained. To combat the weakness, we propose a transformation approach for achievable DoF region analysis, which transforms the decoding condition of transmission scheme w.r.t. phase duration into the achievable DoF region w.r.t. achievable DoF, through achievable DoF tuple expression connecting phase duration and achievable DoF. Via this transformation approach, we can analyze the achievable DoF region of transmission scheme systematically, even if the DoF outer region is not achieved. 
 \end{itemize}

 \subsection{Notations} 
  The scalar, vector, and matrix are denoted by $h,\textbf{h}$, and $\textbf{H}$, respectively. $(\cdot)'$ and $(\cdot)^H$ denote transpose and conjugate-transpose, respectively. $\mathbb{R}_+^n$ denotes a tuple with $n$ non-negative real numbers. The  $\underline{\textbf{h}}$ or $\underline{\textbf{H}}$ is comprised of partial rows of $\textbf{h}$ or $\textbf{H}$. The convex hull of set $S$ is denoted by $\text{Conv}\,S$. The convex hull of a finite set is the set of all convex combinations of its points. The block-diagonal matrix with blocks $\textbf{A}$ and $\textbf{B}$ is denoted by
\begin{equation} 
\text{blkdiag}\{\textbf{A}, \textbf{B}\} =
\begin{bmatrix}
\textbf{A} & \textbf{0} \\
\textbf{0} & \textbf{B}
\end{bmatrix}.
\end{equation}

\section{System Model}

\begin{figure}
	\begin{center} \label{Fig1}
		\begin{tikzpicture}[scale = 0.62]
			\node at (-1.5,2) {$T$};
			\node at (5.5,-1) {$R_3$};
			\node at (5.5,2) {$R_2$};
			\node at (5.5,5) {$R_1$};
			
			\node at (2.25,3.6) {$\textbf{H}_1[t]$};
			\node at (2.25,2) {$\textbf{H}_2[t]$};
			\node at (2.25,0.25) {$\textbf{H}_3[t]$};
			
			\node at (-1,4.5) {$\textbf{H}_i[t-\tau], \tau \in \{1,2,\cdots\}$};
			\node at (-2.5,3.5) {$i =1,2,3$};

			\draw [line width=0.55pt,-latex, dashed] (-1.5,6.5) --  (-1.5,5);
			\draw [line width=0.55pt, dashed] (0,6.5) --  (-1.5,6.5);
			\draw [line width=0.55pt, dashed] (0,6.1) rectangle (3,6.9);
			\draw [line width=0.55pt,-latex, dashed] (5.5,6.5) --  (3,6.5);
			\draw [line width=0.55pt, dashed] (5.5,6) --  (5.5,6.5);
			\node at (1.5,6.45) {$\text{Delay}$};

			\draw [line width=0.65pt] (0,2)--  (1.75,0.7);
			\draw [line width=0.65pt,-latex] (2.8,-0.1)--  (4,-1);
			
			\draw [line width=0.55pt] (0,2)--  (1.65,2);
			\draw [line width=0.55pt,-latex] (2.9,2) -- (4,2);
			
			\draw [line width=0.65pt,-latex] (2.75,4) -- (4,5);
			\draw [line width=0.65pt] (0,2)--  (1.75,3.3);
			
			\draw [line width=0.5pt] (-1,2.7) -- (-0.5,2.7) -- (-0.3,3) -- (-0.3,2.4) -- (-0.5,2.7);
			\node at (-0.7,2.1) {$\vdots$};
			\draw [line width=0.5pt] (-1,1.3) -- (-0.5,1.3) -- (-0.3,1) -- (-0.3,1.6) -- (-0.5,1.3);
			\node at (-0.2,0.25) {$M$};
			
			\node at (-1.5,0.25) {$\textbf{x}[t]$};
			
			\draw [line width=0.5pt]  (4.2,5.65) -- (4.5,5.45) -- (4.2,5.25) -- (4.2,5.65);
			\draw [line width=0.5pt]  (4.5,5.45) -- (5,5.45);
			\node at (4.6,5.1) {$\vdots$};
			\draw [line width=0.5pt]  (4.2,4.7) -- (4.5,4.5) -- (4.2,4.3) -- (4.2,4.7);
			\draw [line width=0.5pt]  (4.5,4.5) -- (5,4.5);
			
			\node at (4.6,3.8) {$N_1$};
			\node at (4.6,0.8) {$N_2$};
			\node at (4.6,-2.2) {$N_3$};
			
			\node at (7,5) {$\textbf{y}_1[t]$};
			\node at (7,2) {$\textbf{y}_2[t]$};
			\node at (7,-1) {$\textbf{y}_3[t]$};
			
			\draw [line width=0.5pt]  (4.2,2.65) -- (4.5,2.45) -- (4.2,2.25) -- (4.2,2.65);
			\draw [line width=0.5pt]  (4.5,2.45) -- (5,2.45);
			\node at (4.6,2.1) {$\vdots$};
			\draw [line width=0.5pt]  (4.2,1.7) -- (4.5,1.5) -- (4.2,1.3) -- (4.2,1.7);
			\draw [line width=0.5pt]  (4.5,1.5) -- (5,1.5);
			
			\draw [line width=0.5pt]  (4.2,-0.35) -- (4.5,-0.55) -- (4.2,-0.75) -- (4.2,-0.35);
			\draw [line width=0.5pt]  (4.5,-0.55) -- (5,-0.55);
			\node at (4.6,-0.9) {$\vdots$};
			\draw [line width=0.5pt]  (4.2,-1.3) -- (4.5,-1.5) -- (4.2,-1.7) -- (4.2,-1.3);
			\draw [line width=0.5pt]  (4.5,-1.5) -- (5,-1.5);

			\draw [line width=0.55pt] (-2,1) rectangle (-1,3);
			\draw [line width=0.55pt]  (5,-1.75) rectangle (6,-0.25);
			\draw [line width=0.55pt]  (5,4.25) rectangle (6,5.75);
			\draw [line width=0.55pt]  (5,1.25) rectangle (6,2.75);
		\end{tikzpicture}
	\end{center}
	\caption{Three-user $(N_1, N_2, N_3, M)$  MIMO broadcast channel with delayed CSIT.}
\end{figure}
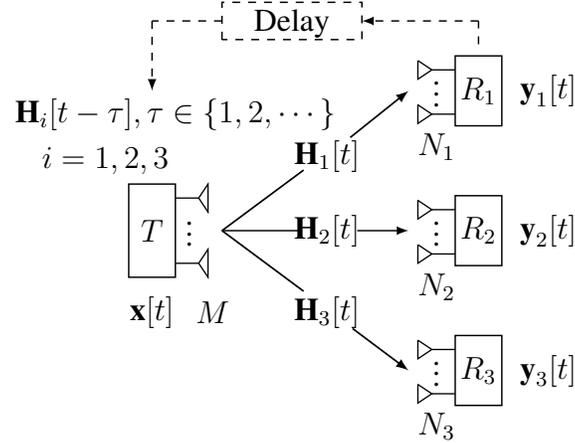

\subsection{Three-user MIMO Broadcast Channel}

The three-user  $(N_1, N_2, N_3, M)$  MIMO broadcast channel with arbitrary antenna configurations at the transmitter and three receivers is depicted in Fig. 1, where the transmitter is equipped with $M$ antennas and receivers 1, 2 and 3 are equipped with $N_1$, $N_2$ and $N_3$ antennas, respectively. Without loss of generality, we assume $N_1 \le N_2 \le N_3$. At the time slot (TS) $t$, the channel state information from the transmitter to the receiver $i = 1,2,3$, is denoted by the matrix $\textbf{H}_i[t] \in \mathbb{C}^{N_i \times M}$, whose elements are i.i.d. across space and time, and drawn from a continuous distribution. The received signal in the TS $t$ at receiver $i$ is expressed as
\begin{equation}  
\textbf{y}_i[t] = \textbf{H}_i[t]\textbf{x}[t] + \textbf{n}_i[t], \qquad i=1,2,3,
\end{equation}
where the input signal at the transmitter is denoted by $\textbf{x}[t]$ and the additive white Gaussian noise (AWGN) at the receiver $i$ is denoted by $\textbf{n}_i[t] \sim {\cal{CN}}(0, \sigma^2)$. Given a maximal average transmit power $P$, $\textbf{x}[t]$ is subject to an average power constraint, i.e., $
\sum_{t=1}^n\textbf{x}[t]^H\textbf{x}[t] \le P
$.

\subsection{Delayed CSIT}

At each TS, each receiver estimates its channel state information matrix and returns it to the transmitter. We assume that the feedback from the receivers to the transmitter is subject to a delay, whose value is not smaller than one TS. Therefore, only delayed channel state information, i.e., $\textbf{H}_i[t-\tau], 1 \le \tau$, is available at the transmitter. Furthermore, we assume the channel state information at receivers (CSIR) is perfect (without delay).

\subsection{DoF Region}

The order-2 message rate tuple $(R_{12}(P), R_{23}(P), R_{13}(P))$ is achievable, if there exists a code such that the probability of decoding error approaches zero when the number of channel uses goes to  infinity.  The channel capacity region of order-2 message ${\cal{C}}_2(P)$ is the region of all achievable rate tuple of order-2 messages satisfying the average power constraint. The  DoF region of order-2 messages is defined as follows: 
 \begin{equation} 
\left\{ 
(d_{12}, d_{23}, d_{13}) \in \mathbb{R}_+^3 \left| 
\begin{split} 
(R_{12}(P), R_{23}(P), R_{13}(P)) \in {\cal{C}}_2(P), \\
d_{j} = \lim_{P \rightarrow {\cal{1}}} \frac{R_{j}(P)}{\log_2 P}, j \in \{12, 23, 13\}.
\end{split}\right.\right\}.
\end{equation}

The order-1 message rate tuple  $(R_{1}(P), R_{2}(P), R_{3}(P))$ is achievable, if there exists a code such that the probability of decoding error approaches zero when the number of channel uses goes to infinity. The channel capacity region ${\cal{C}}_1(P)$ is the region of all achievable rate tuple satisfying the average power constraint. The DoF region of order-2 messages is defined as follows: 
\begin{equation} 
 \left\{ 
(d_{1}, d_{2}, d_{3}) \in \mathbb{R}_+^3 \left| 
\begin{split} 
(R_{1}(P), R_{2}(P), R_{3}(P)) \in {\cal{C}}_1(P), \\
d_{j} = \lim_{P \rightarrow {\cal{1}}} \frac{R_{j}(P)}{\log_2 P}, j \in \{1, 2, 3\}.
\end{split}\right.\right\}.
\end{equation}

\section{DoF Region of Order-2 Messages}
In this section, we first present Lemmas 1-3, and then put forward Theorem 1, where Lemmas 1 and 2 are used as achievability proof of Theorem 1, and Lemma 3 is used to prove the converse  of Theorem 1.

\textbf{Lemma 1}:  If  $M \le N_2$, for the three-user MIMO broadcast channel with delayed CSIT, the DoF region of order-2 messages is given by
\begin{equation} \label{C1}
	{\cal{D}}_2^\text{ach.} =    \left\{
	(d_{12}, d_{23}, d_{13}) \in \mathbb{R}^3_+  \left|
	\begin{split}
		\frac{d_{12} + d_{13}}{\min\{M,N_1\}} + \frac{d_{23}}{M} \le 1
	\end{split}
	\right\}. \right.
\end{equation}
\begin{IEEEproof}
	Please refer to the first part of sub-section B.
\end{IEEEproof}

\textbf{Lemma 2}: If $N_2 < M$, for the three-user MIMO broadcast channel with delayed CSIT, the DoF region of order-2 messages is given by
\begin{equation} \label{C3} 
	{\cal{D}}_2^\text{ach.} =    \left\{
	(d_{12}, d_{23}, d_{13}) \in \mathbb{R}^3_+  \left|
	\begin{split}
		&\frac{d_{12} + d_{13}}{N_1} + \frac{d_{23}}{\min\{M,N_1+N_2\}} \le 1, \\
		&\frac{d_{12} + d_{23}}{N_2} + \frac{d_{13}}{\min\{M,N_1+N_2\}} \le 1, \\
		&\frac{d_{13} + d_{23}}{N_3} + \frac{d_{12}}{\min\{M,N_1+N_3\}} \le 1, (\text{if $N_3 < M$})
	\end{split}  
	\right\} \right., 
\end{equation} 
where the third inequality exists if $N_3 < M$ holds.
\begin{IEEEproof}
	Please refer to the second part of sub-section B.
\end{IEEEproof}

\textbf{Lemma 3}: For order-2 messages, the DoF region of  three-user MIMO broadcast channel with no CSIT is given by
	\begin{equation}
		{\cal{D}}_2^{\text{No}} = \left\{ (d_{12}, d_{23}, d_{13}) \in \mathbb{R}^3_+  \left|\right. \frac{d_{12}+d_{13}}{\min\{M,N_1\}} +  \frac{ d_{23}}{\min\{M,N_2\}} \le 1  \right\}. \label{Col}
\end{equation}
\begin{IEEEproof}
	Please refer  to Appendix A.
\end{IEEEproof}

\textbf{Theorem 1}: For the three-user MIMO broadcast channel with delayed CSIT, the DoF region of order-2 messages is given by
\begin{equation} \label{DoFRegion}
{\cal{D}}_2 =    \left\{
(d_{12}, d_{23}, d_{13}) \in \mathbb{R}^3_+  \left|
\begin{split}
\frac{d_{12} + d_{13}}{\min\{M,N_1\}} + \frac{d_{23}}{\min\{M,N_1 + N_2\}} \le 1,  \\
\frac{d_{12} + d_{23}}{\min\{M,N_2\}} + \frac{d_{13}}{\min\{M,N_1 + N_2\}} \le 1, \\
\frac{d_{13} + d_{23}}{\min\{M,N_3\}} + \frac{d_{12}}{\min\{M,N_1 + N_3\}} \le 1. 
\end{split}
\right\}. \right.
\end{equation}
\begin{IEEEproof} 
The converse proof is provided in sub-section A and the achievability proof is provided in Lemmas 1 and 2. 
\end{IEEEproof}

\subsection{Converse Proof}
To show the converse, we follow the idea in \cite{4}, where similar arguments were used to prove the DoF outer region of order-$j=1,2,\cdots,K$ messages
 in $K$-user MISO broadcast channel with delayed CSIT. To begin with, a genie creates a physically degraded broadcast channel by providing the output of receiver $j$ to receiver $j+1, \cdots, 3$. Then, according to \cite{50,51}, the delayed feedback will not change the capacity region of the physically degraded broadcast channel with no feedback, thus the DoF region of order-2 messages with delayed CSIT is equal to the DoF region of order-2 messages with no CSIT. Finally, by Lemma 3, we conclude that 
 \begin{equation}
 \frac{d_{12} + d_{13}}{\min\{M,N_1\}} + \frac{d_{23}}{\min\{M,N_1 + N_2\}} \le 1,
 \end{equation} 
 for the physically degraded broadcast channel with delayed CSIT. Permuting all possible receiver indexes, we can obtain the following DoF outer region: 
 \begin{equation} \label{DoFRegion1}
 {\cal{D}}_2^\text{outer} =    \left\{
 (d_{12}, d_{23}, d_{13}) \in \mathbb{R}^3_+  \left|
 \begin{split}
 \frac{d_{12} + d_{13}}{\min\{M,N_1\}} + \frac{d_{23}}{\min\{M,N_1 + N_2\}} \le 1, \\
 \frac{d_{12} + d_{13}}{\min\{M,N_1\}} + \frac{d_{23}}{\min\{M,N_1 + N_3\}} \le 1, \\
 \frac{d_{12} + d_{23}}{\min\{M,N_2\}} + \frac{d_{13}}{\min\{M,N_1 + N_2\}} \le 1,\\
 \frac{d_{12} + d_{23}}{\min\{M,N_2\}} + \frac{d_{13}}{\min\{M,N_2 + N_3\}} \le 1, \\
 \frac{d_{13} + d_{23}}{\min\{M,N_3\}} + \frac{d_{12}}{\min\{M,N_1 + N_3\}} \le 1,\\
 \frac{d_{13} + d_{23}}{\min\{M,N_3\}} + \frac{d_{12}}{\min\{M,N_2 + N_3\}} \le 1.
 \end{split}
 \right\}. \right.
 \end{equation}
Furthermore, we notice that there are redundant inequalities in \eqref{DoFRegion1}, i.e.,
\begin{subequations}
    \begin{eqnarray}
    \frac{d_{12} + d_{13}}{\min\{M,N_1\}} + \frac{d_{23}}{\min\{M,N_1 + N_3\}} \le 1, \label{RE1} \\
    \frac{d_{12} + d_{23}}{\min\{M,N_2\}} + \frac{d_{13}}{\min\{M,N_2 + N_3\}} \le 1, \label{RE2} \\
    \frac{d_{13} + d_{23}}{\min\{M,N_3\}} + \frac{d_{12}}{\min\{M,N_2 + N_3\}} \le 1, \label{RE3}
    \end{eqnarray}
\end{subequations} 
which are dominated by the remaining inequalities in \eqref{DoFRegion1}. Hence, eliminating \eqref{RE1}-\eqref{RE3} from \eqref{DoFRegion1}, we obtain the converse of \eqref{DoFRegion}. This completes the proof.

 \subsection{Achievability Proof}

 \subsubsection{Proof of Lemma 1}
 
To achieve corner points of \eqref{C1}, we transmit $\min\{M,N_1\}$ order-2 symbols for receivers 1 and 2, $M$ order-2 symbols for receivers 2 and 3, or $\min\{M,N_1\}$ order-2 symbols for receivers 1 and 3, in each TS. The entire region can be attained through time-sharing of the schemes used in achieving the corner points. This completes the proof.

 \subsubsection{Proof of Lemma 2}
    
 The sketch of the proof is given as follows: First, we design a general two-phase transmission scheme with undetermined number of transmit antennas, i.e., $B_1$, $B_2$, and $B_3$, and undetermined phase duration, i.e., $T_{12}$, $T_{23}$, $T_{13}$, and $T$. For the scheme, all order-2 symbols are transmitted in Phase-I and the order-3 symbols that assist the decoding of order-2 symbols are transmitted in Phase-II. Then, we derive the decoding condition of transmission scheme. Finally, we are able to prove the Lemma 2 by transforming the decoding condition and assigning the specific number of transmit antennas.

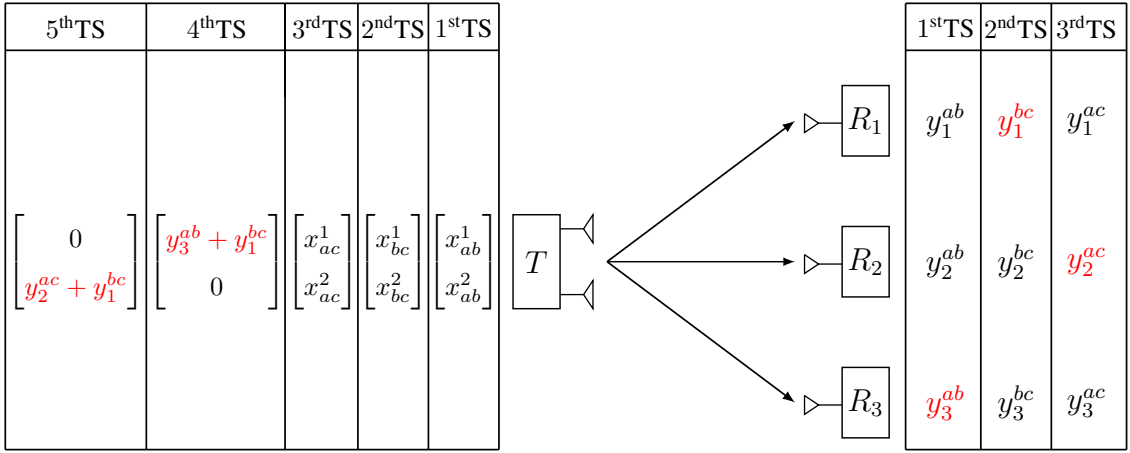
\begin{figure}[!t]
	\begin{center} \label{Fig2}
		\begin{tikzpicture}[scale = 0.625]
			\node at (-1.7,2) {$T$};
			\node at (5.3,-1) {$R_3$};
			\node at (5.3,2) {$R_2$};
			\node at (5.3,5) {$R_1$};

		 \draw [line width=0.65pt ] (-13,7.5)--  (-2.5,7.5);

		 \draw [line width=0.65pt] (-7.05,7.5) -- (-7.05,5.5);

			\draw [line width=0.65pt ] (-13,7.5) -- (-13,-2);
			\draw [line width=0.65pt ] (-13,6.5)--  (-2.5,6.5);
			\draw [line width=0.65pt ] (-7.05,7.5) -- (-7.05,-2);
		
			\draw [line width=0.65pt] (-10,7.5) -- (-10,-2);
			\draw [line width=0.65pt] (-5.5,7.5) -- (-5.5,-2);
			\draw [line width=0.65pt] (-4,7.5) -- (-4,-2);
			\draw [line width=0.65pt ] (-2.5,7.5) -- (-2.5,-2);
			
			\draw [line width=0.65pt] (-13,-2) -- (-2.5,-2);
			
			\draw [line width=0.65pt] (6.15,7.5)--  (10.85,7.5);
			\draw [line width=0.65pt] (6.15,6.5)--  (10.85,6.5);
			\draw [line width=0.65pt] (6.15,-2)--  (10.85,-2);
			
			\draw [line width=0.65pt] (6.15,7.5) -- (6.15,-2);
			\draw [line width=0.65pt] (7.75,7.5) -- (7.75,-2);
			\draw [line width=0.65pt] (9.25,7.5) -- (9.25,-2);
			\draw [line width=0.65pt] (10.85,7.5) -- (10.85,-2);

			\node at (-8.5,7) {{\small $4^\text{th} \text{TS}$}};
			\node at (-11.5,7) {{\small $5^\text{th} \text{TS}$}};
			\node at (-6.25,7) {{\small $3^\text{rd} \text{TS}$}};
			\node at (-4.75,7) {{\small $2^\text{nd} \text{TS}$}};
			\node at (-3.25,7) {{\small $1^\text{st} \text{TS}$}};
			
			\node at (7,7) {{\small $1^\text{st} \text{TS}$}};
			\node at (8.5,7) {{\small $2^\text{nd} \text{TS}$}};
			\node at (10,7) {{\small $3^\text{rd} \text{TS}$}};

			\node at (-3.25,2) {{\small$\begin{bmatrix}
					x_{ab}^1 \\
					x_{ab}^2
			\end{bmatrix}$}};
			
			\node at (-4.75,2) {{\small$\begin{bmatrix}
						x_{bc}^1 \\
						x_{bc}^2
					\end{bmatrix}$}};
			
			\node at (-6.25,2) {{\small$\begin{bmatrix}
						x_{ac}^1 \\
						x_{ac}^2
					\end{bmatrix}$}};
			
				\node at (-8.5,2) {{\small$\begin{bmatrix} 
							\textcolor{red}{y_3^{ab} +  y_1^{bc}}\\ 
							0 
						\end{bmatrix}$}};
					
					\node at (-11.5,2) {{\small$\begin{bmatrix} 
								0\\ 
								\textcolor{red}{y_2^{ac} +  y_1^{bc}} 
							\end{bmatrix}$}};	
			
			\draw [line width=0.65pt,-latex] (-0.2,2)--  (3.8,-1);

			\draw [line width=0.55pt,-latex] (-0.2,2) -- (3.8,2);
			
			\draw [line width=0.65pt,-latex] (-0.2,2) -- (3.8,5);

			\draw [line width=0.5pt] (-1.2,2.7) -- (-0.7,2.7) -- (-0.5,3) -- (-0.5,2.4) -- (-0.7,2.7);
	 
			\draw [line width=0.5pt] (-1.2,1.3) -- (-0.7,1.3) -- (-0.5,1) -- (-0.5,1.6) -- (-0.7,1.3);

		\draw [line width=0.5pt]  (4,5.15) -- (4.3,4.95) -- (4,4.75) -- (4,5.15);
		\draw [line width=0.5pt]  (4.3,4.95) -- (4.8,4.95);

			\node at (7,5) {$y_1^{ab}$};
			\node at (7,2) {$y_2^{ab}$};
			\node at (7,-1) {\textcolor{red}{$y_3^{ab}$}};
				\node at (8.5,5) {\textcolor{red}{$y_1^{bc}$}};
			\node at (8.5,2) {$y_2^{bc}$};
			\node at (8.5,-1) {$y_3^{bc}$};
			\node at (10,5) {$y_1^{ac}$};
			\node at (10,2) {\textcolor{red}{$y_2^{ac}$}};
			\node at (10,-1) {$y_3^{ac}$};
			
			\draw [line width=0.5pt]  (4,2.15) -- (4.3,1.95) -- (4,1.75) -- (4,2.15);
			\draw [line width=0.5pt]  (4.3,1.95) -- (4.8,1.95);

			\draw [line width=0.5pt]  (4,-0.85) -- (4.3,-1.05) -- (4,-1.25) -- (4,-0.85);
			\draw [line width=0.5pt]  (4.3,-1.05) -- (4.8,-1.05);

			\draw [line width=0.55pt] (-2.2,1) rectangle (-1.2,3);
			\draw [line width=0.55pt]  (4.8,-1.75) rectangle (5.8,-0.25);
			\draw [line width=0.55pt]  (4.8,4.25) rectangle (5.8,5.75);
			\draw [line width=0.55pt]  (4.8,1.25) rectangle (5.8,2.75);
		\end{tikzpicture}
	\end{center}
	\caption{Illustration of the exemplified transmission scheme when $M=2$ and $N_1=N_2=N_3=1$, where received signals used as order-2 symbols are colored by red.} \label{II1}
\end{figure}

To begin with, we provide an exemplified transmission scheme when $M=2$ and $N_1=N_2=N_3=1$, as illustrated in Fig. \ref{II1}. In the $1^\text{st}$ TS of Phase-I, two order-2 symbols $x_{ab}^1$ and $x_{ab}^2$ desired by receivers 1 and 2 are transmitted with two  antennas. The received signals are given by 
\begin{equation}
 y_i^{ab}= 	\textbf{h}_i[1] \begin{bmatrix}
		x_{ab}^1 \\
		x_{ab}^2
	\end{bmatrix}, i=1,2,3.   
\end{equation} 
In the $2^\text{nd}$ TS of Phase-I, two order-2 symbols $x_{bc}^1$ and $x_{bc}^2$ desired by receivers 2 and 3 are transmitted with two antennas. The received signals are given by 
\begin{equation} 
  y_i^{bc}= 	\textbf{h}_i[2] \begin{bmatrix}
		x_{bc}^1 \\
		x_{bc}^2
	\end{bmatrix}, i=1,2, 3.
\end{equation}	 
In the $3^\text{rd}$ TS of Phase-I, two order-2 symbols $x_{ac}^1$ and $x_{ac}^2$ desired by receivers 1 and 3 are transmitted with two antennas. The received signals are given by 
\begin{equation}
 y_i^{ac}= 	\textbf{h}_i[3] \begin{bmatrix}
		x_{ac}^1 \\
		x_{ac}^2
	\end{bmatrix}, 
 i =1,2, 3.  
\end{equation}	 
After transmission, in order to decode $x_{ab}^1$ and $x_{ab}^2$, receivers 1 and 2 still need 1 equation. In order to decode $x_{bc}^1$ and $x_{bc}^2$, receivers 2 and 3 still need 1  equation. In order to decode $x_{ac}^1$ and $x_{ac}^2$, receivers 1 and 3 still need 1  equation. Thus, to assist the decoding of order-2 symbols, we design the following order-3 symbols:
\begin{equation}
		\textbf{x}_{abc}  
	=  \begin{bmatrix} 
		y_3^{ab} +  y_1^{bc}\\ 
		y_2^{ac} +  y_1^{bc} 
	\end{bmatrix} \in \mathbb{C}^{2},
\end{equation}
In Phase-II, $y_3^{ab} +  y_1^{bc}$ and $y_2^{ac} +  y_1^{bc}$ are transmitted with 1 TS, respectively, so that they can be decoded at receivers 1 and 2 immediately. To supplement the lacking equations, the receiver 1 can obtain $y_3^{ab}$ by cancellation $y_3^{ab} +  y_1^{bc} - y_1^{bc}$ and $y_2^{ac}$ by cancellation $y_2^{ac} +  y_1^{bc} - y_1^{bc}$, the receiver 2 can obtain $y_1^{bc}$ by cancellation $y_2^{ac} +  y_1^{bc} - y_2^{ac}$ and $y_3^{ab}$ by $y_3^{ab} +  y_1^{bc} - y_1^{bc}$, and receiver 3 can obtain $y_1^{bc}$ by cancellation $y_3^{ab} +  y_1^{bc} - y_3^{ab}$ and $y_2^{ac}$ by cancellation $y_2^{ac} +  y_1^{bc} - y_1^{bc}$. Hence, all order-2 symbols can be decoded. The idea of this specific scheme is extended to the general transmission scheme, which is referred to Appendix B.

For the general transmission scheme, in order to decode the order-2 symbol $\textbf{x}_{ab}$, receiver 1 needs additional $T_{12}(B_1-N_1)$ equations, and receiver 2 needs additional $T_{12}(B_1-N_2)$ equations. In order to decode the order-2 symbol $\textbf{x}_{bc}$, receiver 2 needs additional $T_{23}(B_2 - N_2)$  equations, receiver 3 needs additional $T_{23}(B_2 - N_3)$. In order to decode the order-2 symbol  $\textbf{x}_{ac}$, receiver 1 needs additional $T_{13}(B_3 - N_1)$ equations, and receiver 3 needs additional $T_{13}(B_3 - N_3)$ equations. The Phase-II spans $T$ TSs. Based on the CSIT of Phase-I, the order-3 symbols are generated at the transmitter by \eqref{O3} in Appendix B to assist the decoding of desired order-2 symbols at receivers 1, 2 and 3. Since the phase duration is undetermined, we need to figure out the feasible phase duration, under which the transmitted order-2 symbols are decodable. In the Phase-II, we provide $T_{12}(B_1 - N_1) + T_{13}(B_3 - N_1)$ equations for receiver 1, $T_{12}(B_1$ $- N_2)$ $+$  $T_{23}(B_2 - N_2)$ equations for receiver 2, and $T_{13}(B_3 - N_3) + T_{23}(B_2 - N_3)$ equations for receiver 3, so that the transmitted order-2 symbols in the Phase-I can be decoded at receivers 1, 2, and 3. The feasible phase duration, $T, T_{12}, T_{23}$, and $T_{13}$ should satisfy the following inequalities:
\begin{subequations}  
	\begin{eqnarray}      
		T_{12}(B_1-N_1) + T_{13}(B_3 - N_1) \le TN_1, \label{E4.1.1}\\
		T_{12}(B_1 - N_2) + T_{23}(B_2 - N_2) \le TN_2, \label{E4.1.2} \\
		T_{13}(B_3 - N_3) + T_{23}(B_2 - N_3) \le TN_3, \label{E4.1.3}
	\end{eqnarray}
\end{subequations} 
where \eqref{E4.1.1} represents the number of lacking equations at receiver 1 should be not more than the number of received equations at receiver 1, \eqref{E4.1.2} represents the number of lacking equations at receiver 2 should be not more than the number of received equations at receiver 2, and \eqref{E4.1.3} represents the number of lacking equations at receiver 3 should be not more than the number of received equations at receiver 3. Consequently, we refer the inequalities \eqref{E4.1.1}-\eqref{E4.1.3} to as the decoding condition of this transmission scheme.

Once the decoding condition is obtained, we are able to analyze the achievable DoF region of transmission scheme by transformation approach, whose procedure is detailed as follows: Adding $(\beta  - T)N_1$, $(\beta  - T)N_2$, and $(\beta  - T)N_3$ at both sides of \eqref{E4.1.1}-\eqref{E4.1.3}, respectively, we have
\begin{subequations}  
	\begin{eqnarray}     
		T_{12}B_1 + T_{13}B_3 + T_{23}N_1 \le     \beta N_1, \label{E4.2.1}\\
		T_{12}B_1 + T_{23}B_2  + T_{13}N_2 \le \beta N_2, \label{E4.2.2}\\
		T_{23}B_2 + T_{13}B_3   + T_{12}N_3 \le  \beta N_3. \label{E4.2.3}
	\end{eqnarray}
\end{subequations}
where $\beta  = T_{12} + T_{23} + T_{13} + T$. Then, dividing both sides of  \eqref{E4.2.1}-\eqref{E4.2.3}  by $\beta N_1$, $\beta N_2$ and $\beta N_3$, respectively, we have
\begin{subequations}  
	\begin{eqnarray}     
		\left(\frac{T_{12}B_1}{\beta } + \frac{T_{13}B_3}{\beta }\right)\frac{1}{N_1} + \frac{T_{23}B_2}{\beta }\frac{1}{B_2}   \le 1, \label{E4.3.1} \\
		\left(\frac{T_{12}B_1}{\beta } + \frac{T_{23}B_2}{\beta }\right)\frac{1}{N_2} + \frac{T_{13}B_3}{\beta }\frac{1}{B_3}   \le 1,  
		\label{E4.3.2} \\
			\left(\frac{T_{23}B_2}{\beta}+\frac{T_{13}B_3}{\beta}\right) \frac{1}{N_3} + \frac{T_{12}B_1}{\beta }\frac{1}{B_1} \le 1.
		\label{E4.3.3}
	\end{eqnarray}
\end{subequations} 
The achievable DoF tuple is expressed as 
\begin{equation}  \label{E4.4}
	(d_{12}, d_{23}, d_{13}) = \left(\frac{T_{12}B_1}{\beta }, \frac{T_{23}B_2}{\beta }, \frac{T_{13}B_3}{\beta }\right),
\end{equation} 
where $0 \le T_{12},T_{23}, T_{13}$ implies $0 \le d_{12},d_{23},d_{13}$. Next, we assign specific value to $B_1,B_2$, and $B_3$ so that \eqref{C3} can be achieved. For $N_2 < M \le N_3$ Case, we can set $B_1 = M$ and $B_2=B_3=\min\{M,N_1+N_2\}$. Due to $\min\{M,N_1+N_2\} \le M$ and $M\le N_3$, \eqref{E4.1.3} always satisfies, which implies \eqref{E4.3.3} does not exist in this case.  For $N_3 < M$ Case, we can set $B_1 = \min\{M,N_1+N_3\}$ and $B_2=B_3=\min\{M,N_1+N_2\}$. In all cases, substituting \eqref{E4.4} into \eqref{E4.3.1}-\eqref{E4.3.2} or \eqref{E4.3.1}-\eqref{E4.3.3}, we have \eqref{C3}. This completes the proof.

\section{Achievable DoF Region of Order-1 Messages}

In this section, we present the achievable DoF region of order-1 messages by dividing all the antenna configurations into four specific cases, and show their achievable DoF region via Theorems 2-4, respectively. Moreover, in Corollary 1, we provide the sufficient condition, under which the proposed achievable DoF region is the DoF region. Before we move on, we present the following antenna configuration condition, which is repeatedly used in Theorems 2 and 3, Corollaries 1 and 2:
 \begin{equation}  
  N_1^2(N_3 - N_1) + N_2^2(N_3 - N_2) \le  N_1N_2(N_1 + N_2 - N_3). \eqno(*)  \nonumber
 \end{equation}
As a remark, it can be verified that $N_1 + N_2 <N_3$ leads to the invalidation of the above condition.

\subsection{Case 1: $M \le \max\{N_1 + N_2, N_3\}$}

In this case, an achievable DoF region of order-1 messages was characterized in Theorem 2 of \cite{9}, which is the DoF region if an antenna configuration condition holds.  The sketch of the proof of Theorem 2 in \cite{9} is illustrated as follows: Firstly, a transmission scheme was designed with undetermined phase duration, where the decoding condition w.r.t. phase duration was derived. Secondly, the achievable DoF tuple was substituted into the strictly positive corner point of the DoF outer region, which was subsequently simplified into the expression w.r.t. phase duration. Finally,  if the simplified expression is equal to the decoding condition, then the achievable DoF region is the same as the DoF outer region. Otherwise, the achievable DoF region of transmission scheme is hard to derive, due to the mismatch of DoF outer region.

\subsection{Case 2: $\max\{N_1 + N_2, N_3\} < M \le N_1 +N_3$}

\textbf{Theorem 2}:   For the three-user MIMO broadcast channel with delayed CSIT, if $\max\{N_1 + N_2, N_3\} < M \le N_1 +N_3$,  the achievable DoF region of order-1 messages is given by
\begin{equation}
{\cal{D}}_1^{\text{ach.}} = \text{Conv}  \left\{{\cal{D}}^1, {\cal{D}}^2, {\cal{D}}^3, P_0 \right\},
\end{equation}
where 
$
{\cal{D}}^1 =\{
(d_{2}, d_{3}) \in \mathbb{R}^2_+  |
\frac{d_2}{N_2} + \frac{d_3}{M} \le 1, 
\frac{d_2}{M} +\frac{d_3}{N_3} \le 1.
\},
$
$
{\cal{D}}^2 = \{
(d_{1}, d_{3}) \in \mathbb{R}^2_+ |
\frac{d_1}{N_1} + \frac{d_3}{M} \le 1, 
\frac{d_1}{M} +\frac{d_3}{N_3} \le 1.
\},
$
$
{\cal{D}}^3 = \{
(d_{1}, d_{2}) \in \mathbb{R}^2_+  |
\frac{d_1}{N_1} + \frac{d_2}{N_1+N_2} \le 1, 
\frac{d_1}{N_1+N_2} +\frac{d_2}{N_2} \le 1.
\},
$
and, if the condition $(*)$ holds, the corner point $P_0$ is the intersection of the following planes:
\begin{subequations}  
    \begin{eqnarray} 
    \frac{d_{1}}{N_1} + \frac{d_{2}}{N_1 +N_2} + \frac{d_{3}}{M} = 1,  \label{T2_1} \\
    \frac{d_1}{N_1 + N_2} + \frac{d_2}{N_2} + \frac{d_3}{M} = 1,  \label{T2_2}\\
    \frac{d_1}{M} + \frac{d_2}{M} +\frac{d_3}{N_3} + (d_1 +d_2 - d_3)\frac{M- N_1 - N_2}{2(N_1 + N_2)M} = 1.  \label{T2_3}
    \end{eqnarray}
\end{subequations}
Otherwise, the corner point $P_0$  is the intersection of following planes:
\begin{eqnarray}
    \eqref{T2_1}, \eqref{T2_2}, \text{and} \nonumber \\
    d_1 +d_2 - d_3 = 0.  
\end{eqnarray}

\begin{IEEEproof}
The achievable DoF region in Theorem 2 can be achieved by the time-sharing (convex combination) of ${\cal{D}}^1, {\cal{D}}^2$, ${\cal{D}}^3$, and corner point $P_0$. The regions ${\cal{D}}^1, {\cal{D}}^2$, and ${\cal{D}}^3$ are equal to setting one coordinate of DoF outer region in \cite{5} to zero. Since ${\cal{D}}^1, {\cal{D}}^2$, and ${\cal{D}}^3$ belong to regions of two-user MIMO broadcast channel with delayed CSIT, all of them can be achieved by the scheme in \cite{5}. We shall show that the corner point $P_0$ is achieved by the following three-phase transmission scheme:  

The sketch of the proposed transmission scheme is given as follows: In Phase-I, all the order-1 symbols are transmitted with the assigned number of transmit antennas. After the Phase-I transmission, the receivers cannot decode the desired symbols immediately, due to the lack of received equations and interference. To provide the lacking equations and remove the interference, order-2 symbols are generated at the transmitter with the CSIT of Phase-I. The transmission of Phase-II and Phase-III are through the proposed two-phase transmission scheme in Appendix B.

To begin with, we provide an exemplified transmission scheme when $M=2$ and $N_1 = N_2 = N_3 = 1$. In the $1^\text{st}$ TS of Phase-I, two order-1 symbols 
$x_a^1$ and $x_a^2$ desired by receiver 1 and two order-1 symbols $x_b^1$ and $x_b^2$ desired by receiver 2 are transmitted with two antennas. The designed transmit signal is given by
\begin{equation}
	\textbf{x}[1] = 
	\begin{bmatrix}
		x_a^1 \\
		x_a^2
	\end{bmatrix} + 
	\begin{bmatrix}
	x_b^1 \\
	x_b^2
\end{bmatrix}.
\end{equation}
The received signals are given by
\begin{equation}
y_i[1] = \textbf{h}_i[1]	\begin{bmatrix}
	x_a^1 \\
	x_a^2
\end{bmatrix} + 	\textbf{h}_i[1]
	\begin{bmatrix}
x_b^1 \\
x_b^2
\end{bmatrix}, \quad i=1,2,3.
\end{equation}
In the $2^\text{nd}$ TS of Phase-I,  two order-1 symbols 
$x_b^3$ and $x_b^4$ desired by receiver 2 and two order-1 symbols $x_c^1$ and $x_c^2$ desired by receiver 3 are transmitted with two antennas. The designed transmit signal is given by
\begin{equation}
	\textbf{x}[2] = 
	\begin{bmatrix}
		x_b^3 \\
		x_b^4
	\end{bmatrix} + 
	\begin{bmatrix}
		x_c^1 \\
		x_c^2
	\end{bmatrix}.
\end{equation}
The received signals are given by
\begin{equation}
y_i[2] = \textbf{h}_i[2]	\begin{bmatrix}
		x_b^3 \\
		x_b^4
	\end{bmatrix} + 	\textbf{h}_i[2]
	\begin{bmatrix}
		x_c^1 \\
		x_c^2
	\end{bmatrix}, \quad i=1,2,3.
\end{equation}
In the $3^\text{rd}$ TS of Phase-I, two order-1 symbols $x_a^3$ and $x_a^4$ desired by receiver 1 and two order-1 symbols $x_c^3$ and $x_c^4$ desired by 3 are transmitted with two antennas. The designed transmit signal is given by 
\begin{equation}
	\textbf{x}[3] = 
	\begin{bmatrix}
		x_a^3 \\
		x_a^4
	\end{bmatrix} + 
	\begin{bmatrix}
		x_c^3 \\
		x_c^4
	\end{bmatrix}.
\end{equation}
The received signals are given by
\begin{equation}
y_i[3] = \textbf{h}_i[3]	\begin{bmatrix}
		x_a^3 \\
		x_a^4
	\end{bmatrix} + 	\textbf{h}_i[3]
	\begin{bmatrix}
		x_c^3 \\
		x_c^4
	\end{bmatrix}, \quad i=1,2,3.
\end{equation}
Due to the interference in received signals and lack of equations, each receiver cannot decode its desired symbols. To assist the decoding, as depicted in Fig. \ref{I1}, the order-2 symbols are designed as follows: If we provide $\textbf{h}_1[1][x_b^1, x_b^2]'$ to receivers 1 and 2, then receiver 1 can acquire 1 equation by cancellation $y_1[1] - \textbf{h}_1[1][x_b^1, x_b^2]'$, receiver 2 can 1 equation directly. Hence, $\textbf{h}_1[1][x_b^1, x_b^2]'$ is an order-2 symbol desired by receivers 1 and 2 and is denoted by $x_{ab}^1$. Likewise, $x_{ab}^2 = \textbf{h}_2[1][x_a^1, x_a^2]'$, $x_{bc}^1 = \textbf{h}_3[2][x_b^3, x_b^4]'$, $x_{bc}^2 = \textbf{h}_2[2][x_c^1, x_c^2]'$, $x_{ac}^1 = \textbf{h}_1[3][x_c^3, x_c^4]'$, and $x_{ac}^2 = \textbf{h}_3[3][x_a^3, x_a^4]'$. These order-2 symbols $x_{ab}^1,x_{ab}^2,x_{bc}^1,x_{bc}^2,x_{ac}^1,x_{ac}^2$ are transmitted via the exemplified transmission scheme given in the proof of Lemma 2. The generalized Phase-I transmission and order-2 symbol generation in the above example are provided in Appendix C. In the following, we elaborate on the general transmission scheme.

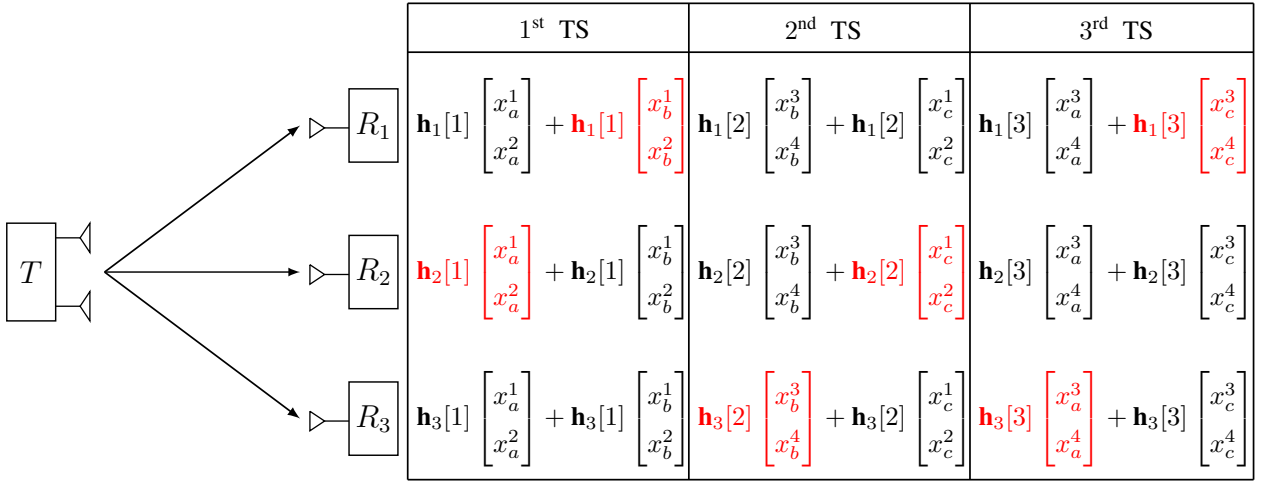
\begin{figure}[!t]
	\begin{center} 
		\begin{tikzpicture}[scale = 0.65]
			\node at (-1.7,2) {$T$};
			\node at (5.3,-1) {$R_3$};
			\node at (5.3,2) {$R_2$};
			\node at (5.3,5) {$R_1$};

			\draw [line width=0.65pt] (6,7.5)--  (23.3,7.5);
			\draw [line width=0.65pt] (6,6.5)--  (23.3,6.5);
			\draw [line width=0.65pt] (6,-2.25)--  (23.3,-2.25);
			
			\draw [line width=0.65pt] (6,7.5) -- (6,-2.25);
			
			\draw [line width=0.65pt] (11.75,7.5) -- (11.75,-2.25);
			
			\draw [line width=0.65pt] (17.5,7.5) -- (17.5,-2.25);
			
			\draw [line width=0.65pt] (23.3,7.5) -- (23.3,-2.25);

			\node at (9,7) {{\small $1^\text{st}  \,\,\, \text{TS}$}};
			
			\node at (14.5,7) {{\small $2^\text{nd} \,\,\, \text{TS}$}};
			
				\node at (20.5,7) {{\small $3^\text{rd}  \,\,\, \text{TS}$}};
			
			\node at (9,5) {{\small$\textbf{h}_1[1]	\begin{bmatrix}
						x_a^1 \\
						x_a^2
					\end{bmatrix} + 	\textcolor{red}{\textbf{h}_1[1]
					\begin{bmatrix}
						x_b^1 \\
						x_b^2
					\end{bmatrix}}$}};
			
			\node at (9,2) {{\small$\textcolor{red}{\textbf{h}_2[1]	\begin{bmatrix}
						x_a^1 \\
						x_a^2
					\end{bmatrix}} + 	\textbf{h}_2[1]
					\begin{bmatrix}
						x_b^1 \\
						x_b^2
					\end{bmatrix}$}};
			
			\node at (9,-1) {{\small$\textbf{h}_3[1]	\begin{bmatrix}
						x_a^1 \\
						x_a^2
					\end{bmatrix} + 	\textbf{h}_3[1]
					\begin{bmatrix}
						x_b^1 \\
						x_b^2
					\end{bmatrix}$}};

			\node at (14.75,5) {{\small$\textbf{h}_1[2]	\begin{bmatrix}
						x_b^3 \\
						x_b^4
					\end{bmatrix} + 	\textbf{h}_1[2]
					\begin{bmatrix}
						x_c^1 \\
						x_c^2
					\end{bmatrix}$}};
			
			\node at (14.75,2) {{\small$\textbf{h}_2[2]	\begin{bmatrix}
						x_b^3 \\
						x_b^4
					\end{bmatrix} + 	\textcolor{red}{\textbf{h}_2[2]
					\begin{bmatrix}
						x_c^1 \\
						x_c^2
					\end{bmatrix}}$}};
			
			\node at (14.75,-1) {{\small$\textcolor{red}{\textbf{h}_3[2]	\begin{bmatrix}
						x_b^3 \\
						x_b^4
					\end{bmatrix}} + 	\textbf{h}_3[2]
					\begin{bmatrix}
						x_c^1 \\
						x_c^2
					\end{bmatrix}$}};
			
			\node at (20.5,5) {{\small$\textbf{h}_1[3]	\begin{bmatrix}
						x_a^3 \\
						x_a^4
					\end{bmatrix} + 	\textcolor{red}{\textbf{h}_1[3]
					\begin{bmatrix}
						x_c^3 \\
						x_c^4
					\end{bmatrix}}$}};
			
			\node at (20.5,2)  {{\small$\textbf{h}_2[3]	\begin{bmatrix}
						x_a^3 \\
						x_a^4
					\end{bmatrix} + 	\textbf{h}_2[3]
					\begin{bmatrix}
						x_c^3 \\
						x_c^4
					\end{bmatrix}$}};
			
			\node at (20.5,-1)  {{\small$\textcolor{red}{\textbf{h}_3[3]	\begin{bmatrix}
						x_a^3 \\
						x_a^4
					\end{bmatrix}} + 	\textbf{h}_3[3]
					\begin{bmatrix}
						x_c^3 \\
						x_c^4
					\end{bmatrix}$}};

			\draw [line width=0.65pt,-latex] (-0.2,2)--  (3.8,-1);

			\draw [line width=0.55pt,-latex] (-0.2,2) -- (3.8,2);
			
			\draw [line width=0.65pt,-latex] (-0.2,2) -- (3.8,5);

			\draw [line width=0.5pt] (-1.2,2.7) -- (-0.7,2.7) -- (-0.5,3) -- (-0.5,2.4) -- (-0.7,2.7);
			
			\draw [line width=0.5pt] (-1.2,1.3) -- (-0.7,1.3) -- (-0.5,1) -- (-0.5,1.6) -- (-0.7,1.3);

			\draw [line width=0.5pt]  (4,5.15) -- (4.3,4.95) -- (4,4.75) -- (4,5.15);
			\draw [line width=0.5pt]  (4.3,4.95) -- (4.8,4.95);

			\draw [line width=0.5pt]  (4,2.15) -- (4.3,1.95) -- (4,1.75) -- (4,2.15);
			\draw [line width=0.5pt]  (4.3,1.95) -- (4.8,1.95);

			\draw [line width=0.5pt]  (4,-0.85) -- (4.3,-1.05) -- (4,-1.25) -- (4,-0.85);
			\draw [line width=0.5pt]  (4.3,-1.05) -- (4.8,-1.05);

			\draw [line width=0.55pt] (-2.2,1) rectangle (-1.2,3);
			\draw [line width=0.55pt]  (4.8,-1.75) rectangle (5.8,-0.25);
			\draw [line width=0.55pt]  (4.8,4.25) rectangle (5.8,5.75);
			\draw [line width=0.55pt]  (4.8,1.25) rectangle (5.8,2.75);
		\end{tikzpicture}
	\end{center}
	\caption{Illustration of the received signals of Phase-I transmission when $M=2$ and $N_1=N_2=N_3=1$, where received signals used as order-2 symbols are colored by red.}
	\label{I1}
\end{figure}

The Phase-I spans $T_1+T_2+T_3$ TSs, where order-1 symbols are transmitted via the strategy in Appendix C with $A_1 = N_1 + N_2$ and $A_2 = A_3 = M$. 
After the transmission of order-1 symbols, receivers 1, 2, and 3 cannot decode the desired symbols, due to the lack of equations and the interference incurred by coded transmission.

The Phase-II spans $T_{12} + T_{23} + T_{13}$ TSs. Utilizing the CSIT of Phase-I, we generate and transmit the order-2 symbol $\textbf{x}_{ab} \in \mathbb{C}^{T_{1}(N_1+N_2)}, \textbf{x}_{bc} \in \mathbb{C}^{T_{2}M}$, and $\textbf{x}_{ac} \in \mathbb{C}^{T_{3}M}$ via the strategy in Appendix C to assist receivers 1, 2 and 3 to decode their desired order-1 symbols. Then, we transmit these order-2 symbols via the strategy in Appendix B, where $B_1 = \min\{M,N_1+N_3\} = M, B_2=B_3= \min\{M,N_1+N_2\} = N_1+N_2$. Therefore, the phase duration for order-2 symbol transmission should satisfy the following equivalence relationship:
\begin{subequations} 
\begin{eqnarray}  
T_{12} = \frac{N_1 + N_2}{M}T_1, \label{R2.1} \\
T_{23} = \frac{M}{N_1 + N_2}T_2, \label{R2.2} \\
T_{13} = \frac{M}{N_1 + N_2}T_3.   \label{R2.3}
\end{eqnarray} 
\end{subequations} 

The Phase-III spans $T$ TSs. Utilizing the CSIT of Phase-II, we generate the order-3 symbol $\textbf{x}_{abc} \in \mathbb{C}^{TN_3}$ via the strategy in Appendix B. In Phase-III, order-3 symbols $\textbf{x}_{abc} \in \mathbb{C}^{TN_3}$ are transmitted  with $N_3$ antennas. The phase duration of Phase-III should be assigned to the value that all receivers can acquire their lacking equations with the same amount of TSs. Thus, according to \eqref{E4.1.1}-\eqref{E4.1.3}, we have 
\begin{subequations} 
    \begin{eqnarray}     
T_{12}(M - N_1) + T_{13}N_2 =     TN_1, \label{Case4.1}\\
T_{12}(M - N_2) + T_{23}N_1 =     TN_2,  \label{Case4.2}\\
(T_{13} + T_{23})(N_1 + N_2 - N_3) =    TN_3, \label{Case4.3}
    \end{eqnarray}
\end{subequations}
 The relationship \eqref{R2.1}-\eqref{R2.3} and linear system \eqref{Case4.1}-\eqref{Case4.3} are the decoding condition. The linear system \eqref{Case4.1}-\eqref{Case4.3} can be solved by the Matlab symbolic calculation, where if the condition $(*)$ holds,
a non-negative solution of the linear system \eqref{Case4.1}-\eqref{Case4.3} can be given by
\begin{subequations} 
    \begin{eqnarray} 
    T_{12} = N_1N_2(N_1 + N_2 - N_3) - N_1^2(N_3 - N_1) 
    - N_2^2(N_3 - N_2), \label{S5.1}\\
    T_{23} = N_2^2(M - N_3) + MN_1(N_3 - N_1), \label{S5.2}\\
    T_{13} = N_1^2(M - N_3) + MN_2(N_3 - N_2), \label{S5.3}\\
    T = (N_1 + N_2 - N_3)(MN_1 - N_1^2 + MN_2 - N_2^2), \label{S5.4}
    \end{eqnarray}
\end{subequations} 
where non-negativity is ensured by  $N_1 \le N_2 \le N_3$ and $N_3 \le N_1+N_2$ implied by the satisfaction of the condition $(*)$, and the other non-negative solutions are the scaling version of  \eqref{S5.1}-\eqref{S5.4}.

Once the decoding condition, i.e., the relationship \eqref{R2.1}-\eqref{R2.3} and linear system \eqref{Case4.1}-\eqref{Case4.3}, is obtained, we are able to derive the achievability of corner point $P_0$ by transformation approach, whose procedure is detailed as follows: Adding $(\alpha-T)N_1$, $(\alpha-T)N_2$ and $(\alpha-T)N_3$ at the both sides of \eqref{Case4.1}-\eqref{Case4.3}, respectively, we have
\begin{subequations} 
    \begin{eqnarray}     
     (T_1 + T_2 + T_3)N_1 + T_{12}M + T_{23}N_1 + T_{13}(N_1 + N_2) = \alpha N_1, \label{R5.1.1}\\
     (T_1 + T_2 + T_3)N_2 + T_{12}M + T_{23}(N_1+N_2) + T_{13}N_2 = \alpha N_2, \label{R5.1.2} \\
    (T_1 + T_2 + T_3)N_3 + T_{12}N_3 + (T_{23} + T_{13})(N_1 + N_2) = \alpha N_3. \label{R5.1.3}
    \end{eqnarray}
\end{subequations}
where $\alpha = T_1 + T_2 + T_3 + T_{12} + T_{23} + T_{13} + T$.
Replacing $T_{12}, T_{23}$, and $T_{13}$ with $T_1, T_2$, and $T_3$ using \eqref{R2.1}-\eqref{R2.3}, and then dividing both sides with $\alpha N_1$, $\alpha N_2$, and $\alpha N_3$, respectively, we have 
\begin{subequations}   
    \begin{eqnarray}  
 \frac{T_1(N_1 + N_2) +  T_3M}{\alpha N_1}  + \frac{T_1(N_1 + N_2) + T_2M}{\alpha (N_1 + N_2)} + \frac{(T_2 + T_3)M}{\alpha M} = 1, \label{R5.3.1} \\
    \frac{T_1(N_1 + N_2) +  T_3M}{\alpha (N_1+N_2)} + \frac{T_1(N_1 + N_2) + T_2M}{\alpha N_2} + \frac{(T_2 + T_3)M}{\alpha M} = 1, \label{R5.3.2}\\
    \frac{T_1(N_1 + N_2) +  T_3M}{\alpha M} + \frac{T_1(N_1 + N_2) + T_2M}{\alpha M} + \frac{(T_2 + T_3)M}{\alpha N_3} + \frac{T_1}{\alpha}\frac{M - N_1 - N_2}{M} = 1. \label{R5.3.3}
    \end{eqnarray}
\end{subequations}  
Since the achievable DoF tuple is expressed as 
\begin{equation} \label{DoF1}
(d_1, d_2, d_3) = \left(\frac{T_1(N_1 + N_2) + T_3M}{\alpha}, \frac{T_1(N_1 + N_2) + T_2M}{\alpha}, \frac{(T_2 + T_3)M}{\alpha}\right),
\end{equation}
we have
\begin{equation} \label{Y2.1}
\frac{T_1}{\alpha} = \frac{d_1 +d_2 - d_3}{2(N_1 + N_2)}.
\end{equation} 
After re-writing \eqref{R5.3.1}-\eqref{R5.3.3} through \eqref{DoF1} and \eqref{Y2.1}, the corner point $P_0$ can be expressed as the intersection of \eqref{T2_1}-\eqref{T2_3}. 
 
If the condition $(*)$ does not hold, we cannot obtain a non-negative solution of the linear system  \eqref{Case4.1}-\eqref{Case4.3}, due to $T_{12} < 0$. Therefore, we shift to solve
\begin{subequations}  
    \begin{eqnarray}  
        TN_1 = T_{12}(M - N_1) + T_{13}N_2,  \\
        TN_2 = T_{12}(M - N_2) + T_{23}N_1,  \\
        T_{12} = 0,
    \end{eqnarray}
\end{subequations} for achieving a corner point. 
A non-negative solution can be given by
$     T_{23} = N_2^2,
    T_{13} = N_1^2,
    T = N_1N_2$.  
Based on \eqref{R2.1}-\eqref{R2.3}, the Phase-I duration can be derived accordingly. Due to the relationship revealed by \eqref{R2.1} and \eqref{Y2.1}, the corner point $P_0$ can be equivalently expressed as the intersection of \eqref{T2_1}, \eqref{T2_2} and $d_1+d_2-d_3=0$. This completes the proof.
\end{IEEEproof}

\textbf{Corollary 1}: If the condition $(*)$ holds with equality, the achievable DoF region in Theorem 2 is the DoF region.  

\begin{IEEEproof}
    According to \cite{5}, a DoF outer region is given by
    \begin{equation} \label{A2} 
    {\cal{D}}_1^{\text{outer}} =    \left\{
    (d_{1}, d_{2}, d_{3}) \in \mathbb{R}^3_+  \left|
    \begin{split}
    \frac{d_{1}}{N_1} + \frac{d_{2}}{N_1 + N_2} + \frac{d_{3}}{M} \le 1, \\
    \frac{d_1}{N_1 + N_2} + \frac{d_2}{N_2} + \frac{d_3}{M} \le 1, \\
    \frac{d_1}{M} + \frac{d_2}{M} +\frac{d_3}{N_3} \le 1.
    \end{split}\right. \right\}.
    \end{equation} 
Due to ${\cal{D}}^i = {\cal{D}}_1^{\text{outer}} \cap d_i = 0, i=1,2,3$, we only need to show the achievability of strictly positive corner point. As shown in the proof of Theorem 2, if the condition $(*)$ holds with equality, we have $T_1 = 0$. This implies $d_1 + d_2 - d_3 = 0$, due to \eqref{Y2.1}. Therefore, in this case, the strictly positive corner point of the DoF outer region is the same as $P_0$. \end{IEEEproof}

\textit{Remark}: Corollary 1 shows the sufficient condition, under which the achievable DoF region in Theorem 2 is the DoF region. Moreover, in Table I, we provide some antenna configurations that satisfy Corollary 1. 

\begin{table} 
	\caption{Some Antenna Configurations Satisfy Corollary 1} 
	\centering 
	\begin{tabular}{|l|c|c|c|c|c|c|c|c|c|c|c|c|c|c|}  
		\hline  
		$N_1$ & 3  & 6 & 7 & 9 & 12 & 13 & 14 & 15 & 18 & 21 & 21 & 21 & 24 & 26\\  
		\hline  
		$N_2$  & 3 & 6 & 14 & 9 & 12 & 39 & 28 & 15 & 18 & 21 & 42 & 84 & 24 & 78 \\  
		\hline
		$N_3$  & 4 & 8 & 15 & 12 & 16 & 40 & 30 & 20 & 24 & 28 & 45 & 85 & 32 & 80 \\  
		\hline
		$M$  & 7 & 13,14  & 22 & 19-21 & 25-28 & 53 & 43,44 & 31-35& 37-42 & 43-49 & 64-66 & 106 & 49-56 & 95-106\\
		\hline  
	\end{tabular}  
\end{table} 

\subsection{Case 3: $N_1 + N_3 < M \le N_2 + N_3$}
 
\textbf{Theorem 3}:  For the three-user MIMO broadcast channel with delayed CSIT, if $N_1 + N_3 < M \le N_2 + N_3$, the achievable DoF region of order-1 messages is given by 
\begin{equation}
{\cal{D}}_1^{\text{ach.}} = \text{Conv} \left\{{\cal{D}}^1, {\cal{D}}^2, {\cal{D}}^3, P_0 \right\},
\end{equation}
where 
$
{\cal{D}}^1 = \{
(d_{2}, d_{3}) \in \mathbb{R}^2_+   |
\frac{d_2}{N_2} + \frac{d_3}{M} \le 1, 
\frac{d_2}{M} +\frac{d_3}{N_3} \le 1.
 \},
$
$
{\cal{D}}^2 =     \{
(d_{1}, d_{3}) \in \mathbb{R}^2_+  | 
\frac{d_1}{N_1 + N_3} + \frac{d_3}{N_3} \le 1$, 
$\frac{d_1}{N_1} +\frac{d_3}{N_1 + N_3} \le 1.
  \},
$
$
{\cal{D}}^3 = \{
(d_{1}, d_{2}) \in \mathbb{R}^2_+ |
\frac{d_1}{N_1} + \frac{d_2}{N_1+N_2} \le 1, 
\frac{d_1}{N_1+N_2} +\frac{d_2}{N_2} \le 1.  \},
$
and, if the condition $(*)$ holds, the corner point $P_0$ is  the intersection of following planes:
\begin{subequations}  
    \begin{eqnarray}
        \frac{d_{1}}{N_1} + \frac{d_{2}}{N_1 +N_2} + \frac{d_{3}}{M} + (d_1 +d_3 - d_2)\frac{M- N_1 - N_3}{2(N_1 + N_3)M} = 1, \label{T3_1}\\
        \frac{d_1}{N_1 + N_2} + \frac{d_2}{N_2} + \frac{d_3}{M} + (d_1 +d_3 - d_2)\frac{M- N_1 - N_3}{2(N_1 + N_3)M} = 1, \label{T3_2} \\
        \frac{d_1}{N_1 +N_3} + \frac{d_2}{M} +\frac{d_3}{N_3} + (d_1 +d_2 - d_3)\frac{M- N_1 - N_2}{2(N_1 + N_2)M} = 1.   \label{T3_3}
    \end{eqnarray}
\end{subequations}
Otherwise, the corner point $P_0$ is the intersection of following planes:
    \begin{eqnarray}
\eqref{T3_1}, \eqref{T3_2}, \text{and} \nonumber \\
    d_1 +d_2 - d_3 = 0.  
    \end{eqnarray} 
 
\begin{IEEEproof}
Similar to Theorem 2, we only need to show that the corner point $P_0$ is achieved by the following three-phase transmission scheme, where the number of used transmit antennas, the number of order-2 and order-3 symbols, and the phase duration are different from the scheme in Theorem 2.

The Phase-I spans $T_1+T_2+T_3$ TSs, where order-1 symbols are transmitted via the strategy in Appendix C with $A_1 = N_1 + N_2$, $A_2 = M$, and $A_3 = N_1+N_3$. After the transmission of order-1 symbols, receivers 1, 2, and 3 cannot decode the desired symbols, due to the lack of equations and the interference incurred by coded transmission.    
    
The Phase-II spans $T_{12} + T_{23} + T_{13}$ TSs. Utilizing the CSIT of Phase-I, we generate the order-2 symbol $\textbf{x}_{ab} \in \mathbb{C}^{T_1(N_1 + N_2)}, \textbf{x}_{bc} \in \mathbb{C}^{T_2M}$ and $\textbf{x}_{ac} \in \mathbb{C}^{T_3(N_1 + N_3)}$ via the strategy  in Appendix C to assist receivers 1, 2, and 3 to decode their desired order-1 symbols. Then, we transmit these order-2 symbols via the strategy in Appendix B, where $B_1 = \min\{M,N_1+N_3\} = N_1+N_3, B_2 = B_3 = \min\{M,N_1+N_2\} = N_1+N_2$. Therefore, the phase duration for order-2 symbol transmission should satisfy the following equivalence relationship:
\begin{subequations}
 \begin{eqnarray} 
T_{12} = \frac{N_1 + N_2}{N_1 + N_3}T_{1}, \label{R3.1} \\ 
T_{23} = \frac{M}{N_1 + N_2}T_{2}, \label{R3.2} \\ 
T_{13} = \frac{N_1 + N_3}{N_1 +N_2}T_{3}. \label{R3.3}
\end{eqnarray}
\end{subequations}

The Phase-III spans $T$ TSs. Utilizing the CSIT of Phase-II, we generate the order-3 symbol $\textbf{x}_{abc} \in \mathbb{C}^{TN_3}$ via the strategy in Appendix B. In Phase-III, all the order-3 symbols are transmitted with $N_3$ antennas. The phase duration of Phase-III should be assigned to the value that all receivers can acquire their lacking equations with the same amount of TSs. Thus, according to \eqref{E4.1.1}-\eqref{E4.1.3}, we have
\begin{subequations}  
    \begin{eqnarray}     
T_{12}N_3 + T_{13}N_2   =  TN_1,  \label{Case5.1}\\
T_{12}(N_1+ N_3 - N_2) + T_{23}N_1  =   TN_2,  \label{Case5.2}\\
(T_{13} + T_{23})(N_1 + N_2 - N_3)  =   TN_3, \label{Case5.3}
    \end{eqnarray}
\end{subequations}
The relationship \eqref{R3.1}-\eqref{R3.3} and linear system \eqref{Case5.1}-\eqref{Case5.3} are the decoding condition. The linear system \eqref{Case5.1}-\eqref{Case5.3} can be solved by the Matlab symbolic calculation, where if the condition $(*)$ holds, a non-negative solution of the linear system  \eqref{Case5.1}-\eqref{Case5.3} can be given by
\begin{subequations} 
    \begin{eqnarray} 
    T_{12} = N_1N_2(N_1 + N_2 - N_3) - N_1^2(N_3 - N_1) 
    - N_2^2(N_3 - N_2), \label{S6.1} \\
    T_{23} = N_2^2(M - N_3) + MN_1(N_3 - N_1), \label{S6.2}\\
    T_{13} = N_1^2(M - N_3) + MN_2(N_3 - N_2), \label{S6.3}\\
    T = (N_1 + N_2 - N_3)N_1N_3 + (N_1 + N_2 - N_3)^2N_2, \label{S6.4}
    \end{eqnarray}
\end{subequations} 
where non-negativity is ensured by $N_1 \le N_2 \le N_3$ and $N_3 \le N_1+N_2$ implied by the satisfaction of the condition $(*)$, and the  other non-negative solutions are scaling version of \eqref{S6.1}-\eqref{S6.4}.

Once the decoding condition, i.e., the relationship \eqref{R3.1}-\eqref{R3.3} and linear system \eqref{Case5.1}-\eqref{Case5.3}, is obtained, we are able to derive the achievability of corner point $P_0$ by transformation approach, whose procedure is detailed as follows: Adding $(\alpha-T)N_1$, $(\alpha-T)N_2$ and $(\alpha-T)N_3$ at the both sides of \eqref{Case5.1}-\eqref{Case5.3} respectively, we have
\begin{subequations}  
    \begin{eqnarray} 
    (T_1 + T_2 + T_3)N_1 + T_{12}(N_1 + N_3) + T_{23}N_1  
    + T_{13}(N_1 + N_2) = \alpha N_1,  \label{R6.1.1}\\
    (T_1 + T_2 + T_3)N_2 + T_{12}(N_1 + N_3)   + T_{23}(N_1+N_2) + T_{13}N_2 = \alpha N_2,  \label{R6.1.2} \\
    (T_1 + T_2 + T_3)N_3 + T_{12}N_3  + (T_{23} + T_{13})(N_1 + N_2) = \alpha N_3. 
    \label{R6.1.3}
    \end{eqnarray}
\end{subequations}
where $\alpha = T_1 + T_2 + T_3 + T_{12} + T_{23} + T_{13} + T$. Replacing $T_{12}, T_{23}$, and $T_{13}$ with $T_1, T_2$, and $T_3$ using \eqref{R3.1}-\eqref{R3.3},  and then dividing both sides with $\alpha N_1$, $\alpha N_2$, and $\alpha N_3$, respectively, we have    
\begin{subequations}
        \begin{eqnarray}
     \frac{T_1(N_1 + N_2) + T_3(N_1+N_3)}{\alpha N_1} + \frac{T_1(N_1 + N_2) + T_2M}{\alpha (N_1 + N_2)}  + \frac{T_2M+ T_3(N_1 + N_3)}{\alpha M} \nonumber \\    
     + \frac{T_3}{\alpha}\frac{M-N_1 - N_3}{M} = 1, \label{R6.3.1} \\
 \frac{T_1(N_1 + N_2) + T_3(N_1+N_3)}{\alpha (N_1 +N_2)} + \frac{T_1(N_1 + N_2) + T_2M}{\alpha N_2}  + \frac{T_2M + T_3(N_1 + N_3)}{\alpha M} \nonumber \\        + \frac{T_3}{\alpha}\frac{M-N_1 - N_3}{M} = 1, \label{R6.3.2} \\
 \frac{T_1(N_1 + N_2) + T_3(N_1+N_3)}{\alpha (N_1 +N_3)} + \frac{T_1(N_1 + N_2) + T_2M}{\alpha M} + \frac{T_2M + T_3(N_1 + N_3)}{\alpha N_3} \nonumber \\    
    + \frac{T_1}{\alpha}\frac{M-N_1 - N_2}{M} = 1. \label{R6.3.3}
    \end{eqnarray}
\end{subequations}
Since the achievable DoF tuple is expressed as 
 \begin{equation}\label{DoF2} 
(d_{1}, d_{2}, d_{3}) =
\left(\frac{T_1(N_1 + N_2) + T_3(N_1 + N_3)}{\alpha}, \frac{T_1(N_1 + N_2) + T_2M}{\alpha},  \frac{T_2M + T_3(N_1 + N_3)}{\alpha}\right),
\end{equation}
we have
\begin{subequations}  
    \begin{eqnarray}
    \frac{T_1}{\alpha} = \frac{d_1 +d_2 - d_3}{2(N_1 + N_2)}, \label{Y3.1} \\ 
    \frac{T_3}{\alpha} = \frac{d_1 +d_3 - d_2}{2(N_1 + N_3)}. \label{Y3.2}
    \end{eqnarray}
\end{subequations}
After re-writing \eqref{R6.3.1}-\eqref{R6.3.3} through \eqref{DoF2}, \eqref{Y3.1} and \eqref{Y3.2}, the corner point $P_0$ can be expressed as an intersection of \eqref{T3_1}-\eqref{T3_3}.

If the condition $(*)$ does not hold, we cannot obtain a non-negative solution of the linear system \eqref{Case5.1}-\eqref{Case5.3}, due to $T_{12} < 0$. Therefore, we shift to solve
\begin{subequations} 
    \begin{eqnarray} 
    TN_1 = T_{12}N_3 + T_{13}N_2, \\
    TN_2 = T_{12}(N_1+ N_3 - N_2) + T_{23}N_1, \\
    T_{12} = 0,
    \end{eqnarray}
\end{subequations}
for achieving a corner point. A non-negative solution can be given by
$   T_{23} = N_2^2$,
    $T_{13} = N_1^2$,  
    $T = N_1N_2 $.
Based on \eqref{R3.1}-\eqref{R3.3}, the Phase-I duration can be derived accordingly. Due to the relationship revealed by \eqref{Y3.1} and \eqref{Y3.2}, the corner point $P_0$ can be equivalently expressed as the intersection of \eqref{T3_1}, \eqref{T3_2} and $d_1 + d_2 - d_3 = 0$. This completes the proof.
\end{IEEEproof}
 
\textbf{Corollary 2}: If the condition $(*)$ holds with equality and $M$ gradually reduces to $N_1+N_3$, the achievable DoF region in Theorem 3 approaches the DoF region.  

\begin{IEEEproof}
    According to \cite{5}, a DoF outer region is given by
    \begin{equation} \label{A3} 
    {\cal{D}}_1^{\text{outer}} \triangleq    \left\{
    (d_{1}, d_{2}, d_{3}) \in \mathbb{R}^3_+  \left|
    \begin{split}
    \frac{d_{1}}{N_1} + \frac{d_{2}}{N_1 + N_2} + \frac{d_{3}}{M} \le 1, \\
    \frac{d_1}{N_1 + N_2} + \frac{d_2}{N_2} +\frac{d_3}{M} \le 1, \\
    \frac{d_1}{N_1 + N_3} + \frac{d_2}{M} +\frac{d_3}{N_3} \le 1, \\
    \frac{d_1}{N_1} + \frac{d_2}{M} + \frac{d_3}{N_1 + N_3} \le 1. 
    \end{split} \right. \right\}.
    \end{equation}
    Due to ${\cal{D}}^i = {\cal{D}}_1^{\text{outer}} \cap d_i = 0, i=1,2,3$, we only need to examine the strictly positive corner point. As shown in the proof of Theorem 3, if the condition $(*)$ holds with equality, we have $T_1 = 0$. This implies $d_1 + d_2 - d_3 = 0$, due to \eqref{Y3.1}. In addition, when $M$  gradually reduces to $N_1+N_3$, $\frac{d_1}{N_1} + \frac{d_2}{M} + \frac{d_3}{N_1 + N_3} \le 1$ is asymptotically dominated by  $\frac{d_{1}}{N_1} + \frac{d_{2}}{N_1 + N_2} + \frac{d_{3}}{M} \le 1$ in \eqref{A3}, 
 meanwhile $(d_1 +d_3 - d_2)\frac{M- N_1 - N_3}{2(N_1 + N_3)M}$ in \eqref{T3_1}-\eqref{T3_2} goes to zero. Therefore, in this case, the corner point $P_0$ approaches the strictly positive corner point of the DoF outer region. 
\end{IEEEproof}

\subsection{Case 4: $N_2 + N_3 < M$}
 \textbf{Theorem 4}: For the three-user MIMO broadcast channel with delayed CSIT, if $N_2 + N_3 < M$, the achievable DoF region of order-1 messages is given by
\begin{equation}
{\cal{D}}_1^{\text{ach.}} = \text{Conv} \left\{{\cal{D}}^1, {\cal{D}}^2, {\cal{D}}^3, P_0 \right\},
\end{equation}
where 
$
{\cal{D}}^{1} = \{
(d_{2}, d_{3}) \in \mathbb{R}^2_+  |
\frac{d_2}{N_2} + \frac{d_3}{N_2 + N_3} \le 1,
\frac{d_2}{N_2+N_3} +\frac{d_3}{N_3} \le 1.
\},
$
$
{\cal{D}}^2 =\{
(d_{1}, d_{3}) \in \mathbb{R}^2_+  |
\frac{d_1}{N_1 + N_3} + \frac{d_3}{N_3} \le 1,
\frac{d_1}{N_1} +\frac{d_3}{N_1 + N_3} \le 1.
\},
$
$
{\cal{D}}^3 = \{
(d_{2}, d_{3}) \in \mathbb{R}^2_+ |
\frac{d_1}{N_1} + \frac{d_2}{N_1+N_2} \le 1,
\frac{d_1}{N_1+N_2} +\frac{d_2}{N_2} \le 1.
\},
$
and the corner point $P_0$ is the intersection of following planes:
\begin{subequations} 
    \begin{eqnarray}
    \frac{d_{1}}{N_1} + \frac{d_{2}}{N_1 +N_2} + \frac{d_{3}}{M} + (d_1 +d_2 - d_3)\frac{MN_2- N_1N_2 - N_2^2}{2(N_1 + N_2)N_1M} = 1, \label{T4.1}\\
    \frac{d_1}{M} + \frac{d_2}{N_2} + \frac{d_3}{N_1 + N_2} + (d_2 +d_3 - d_1)\frac{MN_1- N_1N_3 - N_2N_3}{2(N_1 + N_2)N_2M} = 1,  \label{T4.2}\\
    \frac{d_1}{N_1 +N_3} + \frac{d_2}{M} +\frac{d_3}{N_3} + (d_1 +d_3 - d_2)\frac{MN_1- N_1N_3 - N_1^2}{2(N_1 + N_3)N_3M} = 1. \label{T4.3} 
    \end{eqnarray}
\end{subequations}  
Meanwhile, the corner point $P_0$ should satisfy 
\begin{subequations} 
    \begin{eqnarray}
0 \le d_1 +d_2 - d_3, \label{T4.4}\\
0 \le d_2 +d_3 - d_1,  \label{T4.5}\\
0 \le d_1 +d_3 - d_2. \label{T4.6} 
    \end{eqnarray}
\end{subequations} 
\textit{Remark}: If antenna configurations are symmetric ($N_1 = N_2 = N_3 = N$), the corner point $P_0$ is given by  
$\left(\frac{4MN}{7M +2N}, \frac{4MN}{7M +2N}, \frac{4MN}{7M +2N}\right)
$. In fact, Theorem 4 generalizes the result in our conference paper \cite{8} for symmetric antenna configurations to the setting of arbitrary antenna configurations.

\begin{IEEEproof}
Similar to Theorem 2, we only need to show that the corner point $P_0$ is achieved by the following three-phase transmission scheme:

The sketch of the proposed transmission scheme is given as follows: In Phase-I, all the order-1 symbols are transmitted with the assigned number of transmit antennas. After the Phase-I transmission, the receivers cannot decode the desired symbols immediately, due to the lack of received equations. To provide the lacking equations, order-2 symbols and order-3 symbols are generated at the transmitter with the CSIT of Phase-I. The transmission of the Phase-II and Phase-III are through the proposed two-phase transmission scheme in Section-III. In the following, we elaborate on the proposed transmission scheme as follows:

The Phase-I spans $T_1+T_2+T_3$ TSs, where order-1 symbols are transmitted via the strategy in Appendix C with $A_1 = A_2 =A_3 = M$. After the transmission of order-1 symbols, receivers 1, 2, and 3 cannot decode the desired symbols, due to the lack of equations and the interference incurred by coded transmission.

The Phase-II spans $T_{12} + T_{23} + T_{13}$ TSs. Utilizing the CSIT of Phase-I, we generate the order-2 symbol $\textbf{x}_{ab} \in \mathbb{C}^{T_1(N_1 + N_2)}, \textbf{x}_{bc} \in \mathbb{C}^{T_2(N_2+N_3)}$ and $\textbf{x}_{ac} \in \mathbb{C}^{T_3(N_1 + N_3)}$ via the strategy in Appendix C for assisting receivers 1, 2, and 3 to decode the desired order-1 symbols. Nevertheless, this is not enough. We further generate additional order-2 symbols and order-3 symbols for completely decoding order-1 symbols. The additional order-2 symbols are given by
\begin{subequations}
    \begin{eqnarray} 
        \underline{\textbf{H}}_3^{a+b}\textbf{x}_a^1  + \underline{\textbf{H}}_1^{b+c}(\textbf{x}_b^2 + \textbf{x}_c^1) \in \mathbb{C}^{T_1(M - N_1 - N_2)}, \label{Y0.1} \\
        \underline{\textbf{H}}_1^{b+c}\textbf{x}_b^2  + \underline{\textbf{H}}_2^{a+c}(\textbf{x}_a^2 + \textbf{x}_c^2) \in \mathbb{C}^{T_2(M- N_2 - N_3)}, \label{Y0.2}\\
        \underline{\textbf{H}}_2^{a+c}\textbf{x}_c^2  + \underline{\textbf{H}}_3^{b+c}(\textbf{x}_a^1 + \textbf{x}_b^1) \in \mathbb{C}^{T_3(M-N_1 - N_3)}.  \label{Y0.3}
    \end{eqnarray}
\end{subequations}
The additional order-3 symbols are given by
\begin{equation} 
    \underline{\textbf{H}}_3^{b+c}(\textbf{x}_a^1 + \textbf{x}_b^1) + \underline{\textbf{H}}_1^{b+c}(\textbf{x}_b^2 + \textbf{x}_c^1)  +
    \underline{\textbf{H}}_2^{a+c}(\textbf{x}_a^2 + \textbf{x}_c^2) 
    \in \mathbb{C}^{\max\{T_1(M - N_1 - N_2), T_2(M - N_2 - N_3), T_3(M - N_1 - N_3)\}}. \label{Y0}
\end{equation}  
Then, we transmit these order-2 symbols via the strategy in Appendix B, where $B_1 = \min\{M,N_1+N_3\} = N_1+N_3, B_2 = B_3 = \min\{M,N_1+N_2\} = N_1+N_2$. Therefore, the phase duration for order-2 symbol transmission should satisfy the following equivalence relationship:
\begin{subequations} \label{Y4}
\begin{eqnarray} 
    T_{12} =  \frac{M}{N_1 + N_3}T_{1}, \label{Y4.1} \\ 
    T_{23} = \frac{M}{N_1 + N_2}T_{2}, \label{Y4.2} \\  
    T_{13} =  \frac{M}{N_1 + N_2}T_{3}.   \label{Y4.3}
\end{eqnarray}
\end{subequations}

The Phase-III spans $T$ TSs. Utilizing the CSIT of Phase-II, we generate the order-3 symbol $\textbf{x}_{abc} \in \mathbb{C}^{TN_3}$ via the strategy in Appendix B. In Phase-III, all the order-3 symbols are transmitted  with $N_3$ antennas. The phase duration of Phase-III should be assigned to the value that all receivers can acquire their lacking equations with the same amount of TSs. Thus, considering \eqref{E4.1.1}-\eqref{E4.1.3} and \eqref{Y0}, we have
\begin{subequations} 
    \begin{eqnarray}
T_{12}N_3 + T_{13}N_2 + T_1(M - N_1 - N_2) = TN_1, \label{ES5.1.1} \\
T_{12}(N_1+ N_3 - N_2) + T_{23}N_1 + T_2(M - N_2 - N_3) = TN_2, \label{ES5.1.2} \\
(T_{13} + T_{23})(N_1 + N_2 - N_3) + T_3(M - N_1 - N_3) = TN_3. 
    \label{ES5.1.3}    
    \end{eqnarray}
\end{subequations}
The relationship  \eqref{Y4.1}-\eqref{Y4.3} and linear system \eqref{ES5.1.1}-\eqref{ES5.1.3}  are decoding condition. 

Once the decoding condition is obtained, we are able to derive the achievability of corner point $P_0$ by  transformation approach, whose procedure is detailed as follows: Adding $(\alpha-T)N_1$, $(\alpha-T)N_2$ and $(\alpha-T)N_3$ at  both sides of \eqref{ES5.1.1}-\eqref{ES5.1.3}, respectively, we have    
\begin{subequations}
    \begin{eqnarray}    
  (T_1 + T_2 + T_3)N_1 + T_{12}(N_1 + N_3) + T_{13}(N_1 + N_2) + T_{23}N_1   +T_1(M - N_1 - N_2) =     \alpha N_1, \label{R7.2.1}\\
 (T_1 + T_2 + T_3)N_2 + T_{12}(N_1 + N_3) + T_{23}(N_1 + N_2) + T_{13}N_2  +T_2(M - N_2 - N_3) = \alpha N_2, \label{R7.2.2} \\
  (T_1 + T_2 + T_3)N_3 + T_{12}N_3 + (T_{23} + T_{13})(N_1 + N_2)   +T_3(M - N_1 - N_3) =     \alpha N_3. \label{R7.2.3}
    \end{eqnarray}
\end{subequations}
where $\alpha = T_1 + T_2 + T_3 + T_{12} + T_{23} + T_{13} + T$. Replacing $T_{12}, T_{23}$, and $T_{13}$ with $T_1, T_2$, and $T_3$ using \eqref{Y4.1}-\eqref{Y4.3}, and then dividing both sides of \eqref{R7.2.1}-\eqref{R7.2.3} with $\alpha N_1$, $\alpha N_2$, and $\alpha N_3$, respectively, we have
\begin{subequations} 
\begin{eqnarray} 
\frac{(T_1 + T_3)M}{\alpha N_1} + \frac{(T_1 + T_2)M}{\alpha (N_1 + N_2)}  + \frac{(T_2 + T_3)M}{\alpha M} + \frac{T_1}{\alpha}\frac{MN_2 - N_1N_2 - N_2^2}{(N_1 + N_2)N_1} = 1, \label{R7.3.1}\\
\frac{(T_1 + T_3)M}{\alpha M} +  \frac{(T_1 + T_2)M}{\alpha N_2} + \frac{(T_2 + T_3)M}{\alpha (N_1 + N_2)}  + \frac{T_2}{\alpha}\frac{MN_1 - N_1N_3 - N_2N_3}{(N_1 + N_2)N_2} = 1,
    \label{R7.3.2} \\
\frac{(T_1 + T_3)M}{\alpha (N_1 + N_3)} + \frac{(T_1 + T_2)M}{\alpha M} + \frac{(T_2 + T_3)M}{\alpha N_3} + \frac{T_3}{\alpha}\frac{MN_1 - N_1N_3 - N_1^2}{(N_1 + N_3)N_3} = 1. \label{R7.3.3}
    \end{eqnarray}\end{subequations}
Since the achievable DoF tuple is expressed as 
\begin{equation}\label{DoF3}  
(d_1, d_2, d_3) = \left(\frac{(T_1 + T_3)M}{\alpha}, \frac{(T_1 + T_2)M}{\alpha}, \frac{(T_2 + T_3)M}{\alpha}\right), 
\end{equation} we have
\begin{subequations}
    \begin{eqnarray}
    \frac{T_1}{\alpha} = \frac{d_1 +d_2 - d_3}{2M}, \label{Y5.1} \\ 
    \frac{T_2}{\alpha} = \frac{d_2 +d_3 - d_1}{2M},  \label{Y5.2} \\
    \frac{T_3}{\alpha} = \frac{d_1 +d_3 - d_2}{2M}. \label{Y5.3}
    \end{eqnarray}
\end{subequations}
After re-writing \eqref{R7.3.1}-\eqref{R6.3.3} through \eqref{DoF3}, \eqref{Y5.1}-\eqref{Y5.3}, the corner point $P_0$ can be expressed as an intersection of  \eqref{T4.1}-\eqref{T4.3}. To ensure non-negative  solution of the linear system \eqref{ES5.1.1}-\eqref{ES5.1.3}, due to \eqref{Y4.1}-\eqref{Y4.3} and \eqref{Y5.1}-\eqref{Y5.3}, we require \eqref{T4.4}-\eqref{T4.6} hold. This completes the proof.
\end{IEEEproof}

\subsection{Compared with DoF Outer Region}

In this section, for order-1 messages, we compare the proposed achievable DoF region with the DoF outer region in \cite{5}. Numerically, we calculate the achievable sum-DoF from the proposed achievable DoF region and the sum-DoF upper bound from the DoF outer region in \cite{5}. The sum-DoF upper bound is defined as maximizing $d_1 + d_2 + d_3$ over the DoF outer region and the achievable sum-DoF is defined as maximizing $d_1 + d_2 + d_3$ over the achievable DoF region. It can be verified that the optimal solution is the strictly positive corner point.
 
 \begin{figure}[!ht]
     \begin{center}
         \includegraphics[width=5in]{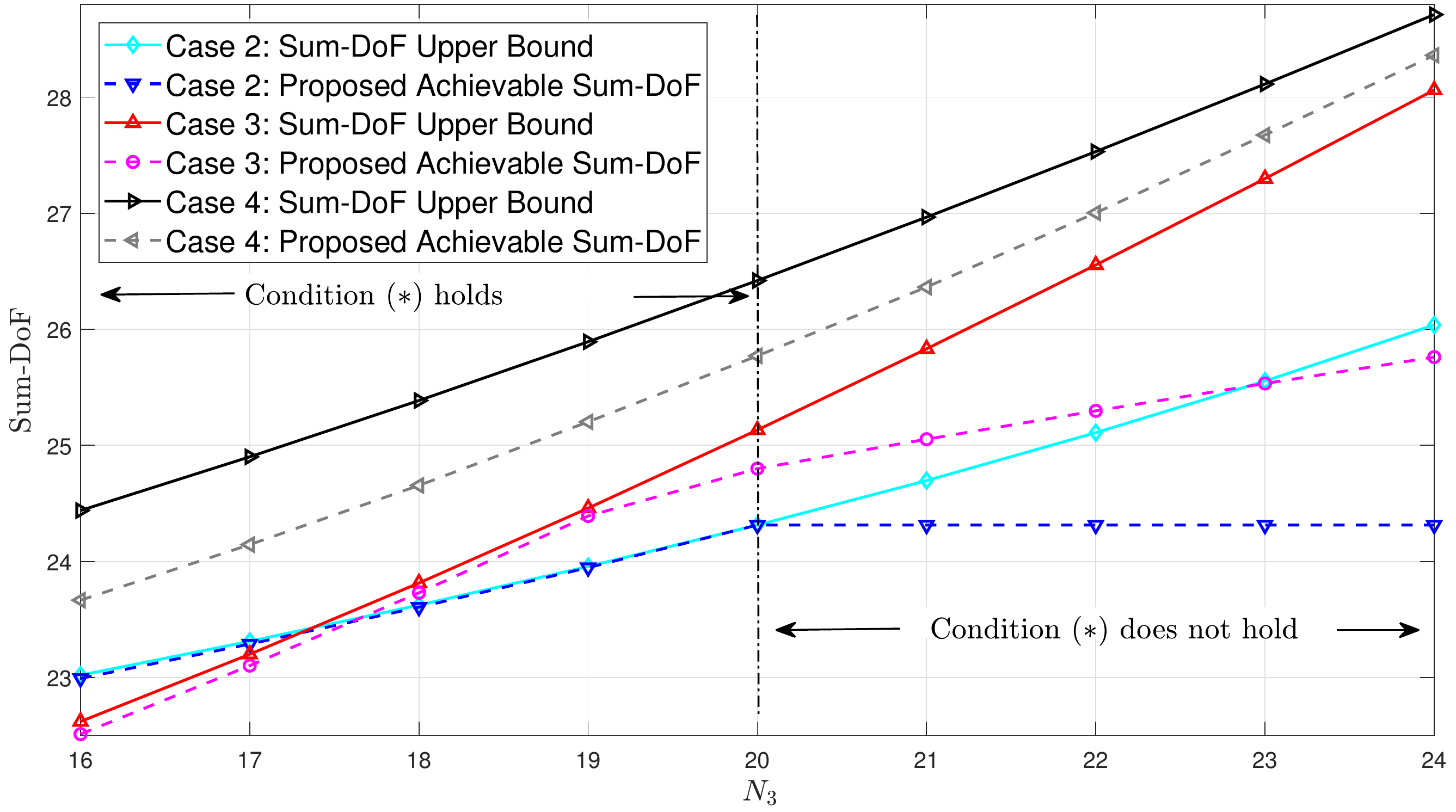}
     \end{center}
     \caption{We set $(N_1, N_2, M) = (15, 15, 31)$ for Case 2, $(N_1, N_2, M) = (14, 15, N_3 + 15)$ for Case 3, and $(N_1, N_2, M) = (15, 15, 40)$ for Case 4.}
     \label{f1}
 \end{figure} 

In Case 2, Fig. \ref{f1} depicts that, if the condition $(*)$ holds, the proposed achievable sum-DoF is extremely close to the sum-DoF upper bound, which implies the satisfactory performance of proposed scheme. If the condition $(*)$ is not satisfied, the DoF gap between the achievable sum-DoF and sum-DoF upper bound is increasingly larger. This is because, in Case 2, the sum-DoF upper bound is related to $N_3$, but the proposed achievable sum-DoF is not affected by $N_3$. In Case 3, Fig. \ref{f1} shows that, if the condition $(*)$ holds, the proposed achievable sum-DoF is close to the sum-DoF upper bound, however, the difference is greater than that of Case 2, and the gap will also grow as $N_3$ increases. In Case 4, which is irrelevant to the condition $(*)$, Fig. \ref{f1} depicts that the gap between sum-DoF upper bound and achievable sum-DoF reduces as  $N_3$ approaches $M$. To sum up, Fig. \ref{f1} shows that the condition $(*)$ is critical because whether it holds or not significantly affects the performance of the proposed design. We can infer from Fig. \ref{f1} that, in Case 4, the larger is the difference between $M$ and $N_2+N_3$, the greater is the performance gap.

\section{Comparison of DoF Regions with Perfect, Delayed, and no CSIT}

In this section, the usefulness of delayed CSIT is verified by comparison of the DoF regions with perfect, delayed, and no CSIT. For order-2 and order-1 messages, we shall show that the DoF region with delayed CSIT are larger than the DoF region with no CSIT if $N_2 < M$, which contains the majority of antenna configurations. While, according to \cite{1,2,3,9} and results of this paper, the DoF region with delayed CSIT are not larger than the one with perfect CSIT. 

\begin{figure}[!ht]
	\begin{minipage}{0.5\linewidth}
		\centering
		\includegraphics[width=1.65in]{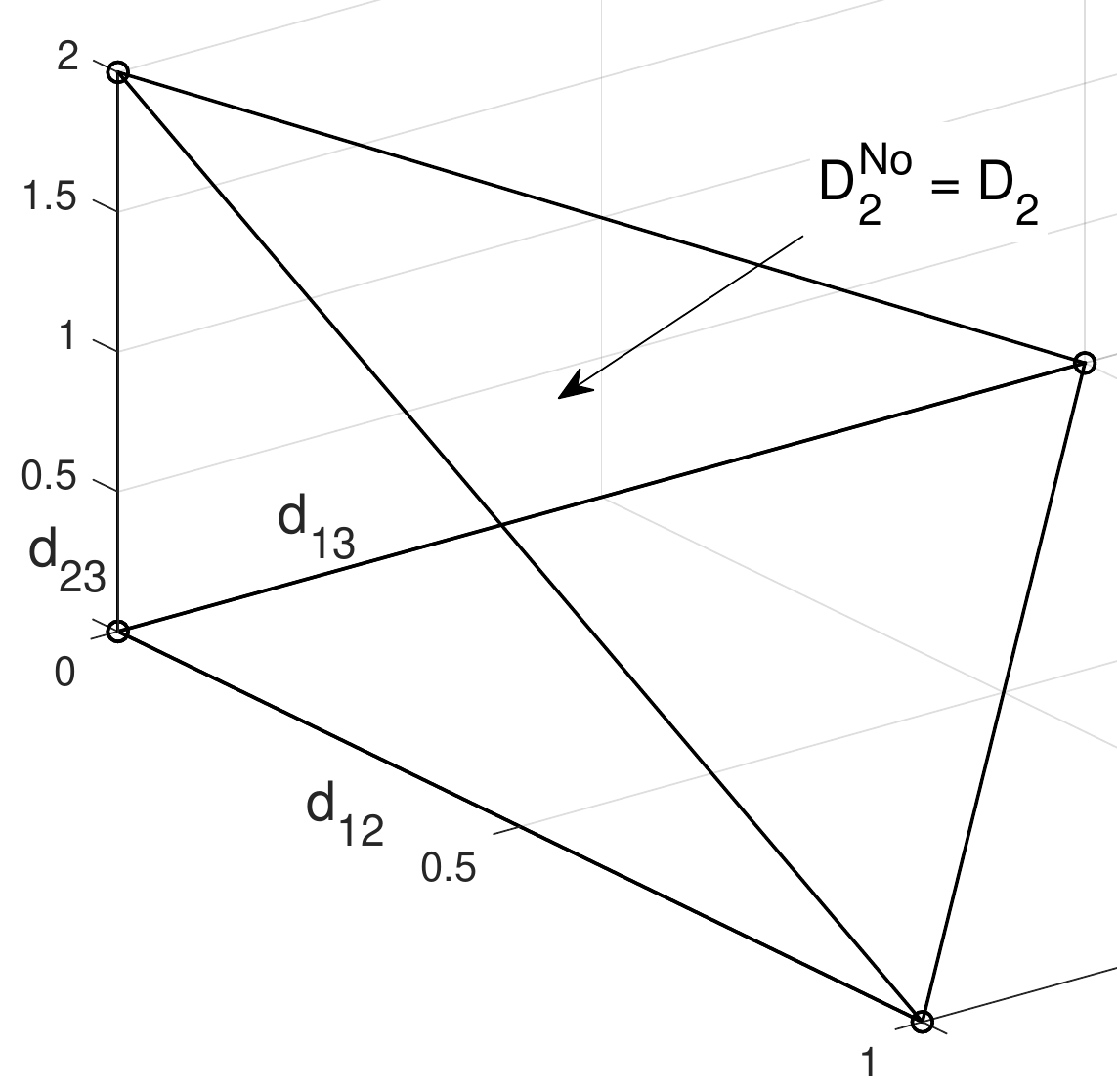}
		\caption{Order-2 messages: $(N_1,N_2,N_3,M) = (1,2,3,2).$}
		\label{fig:side:a}
	\end{minipage}%
	\begin{minipage}{0.5\linewidth}
		\centering
		\includegraphics[width=1.32in]{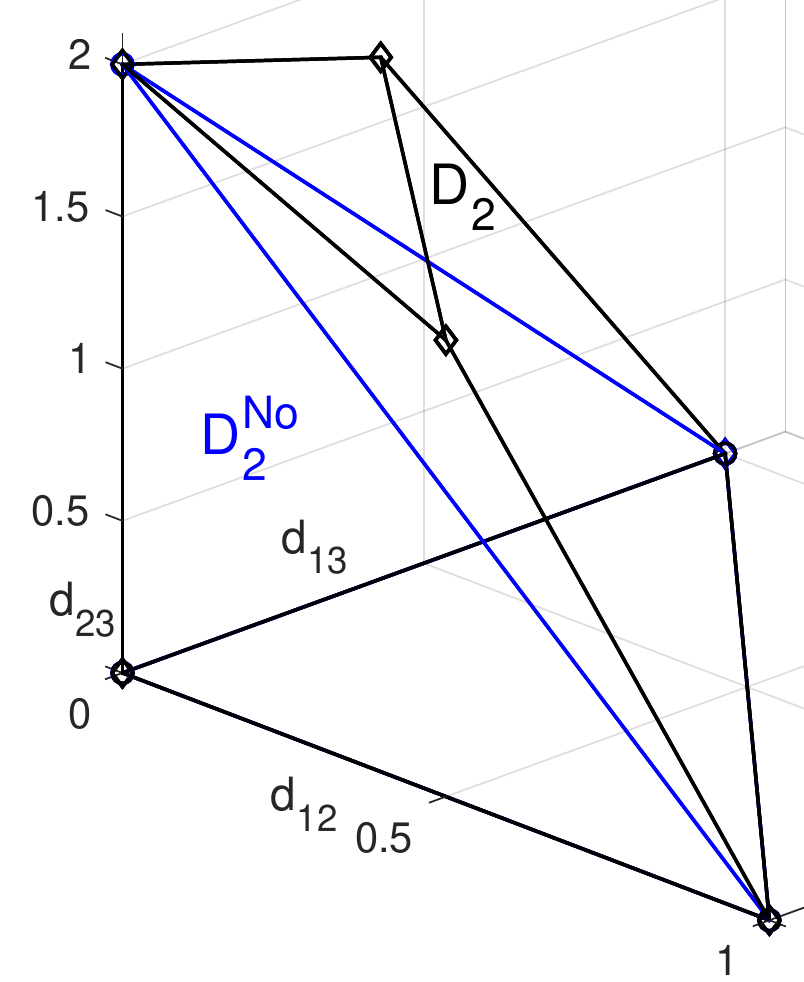}
		\caption{Order-2 messages: $(N_1,N_2,N_3,M) = (1,2,3,3).$}
		\label{fig:side:b}
	\end{minipage} 
	\begin{minipage}{0.5\linewidth}
		\centering
		\includegraphics[width=1.65in]{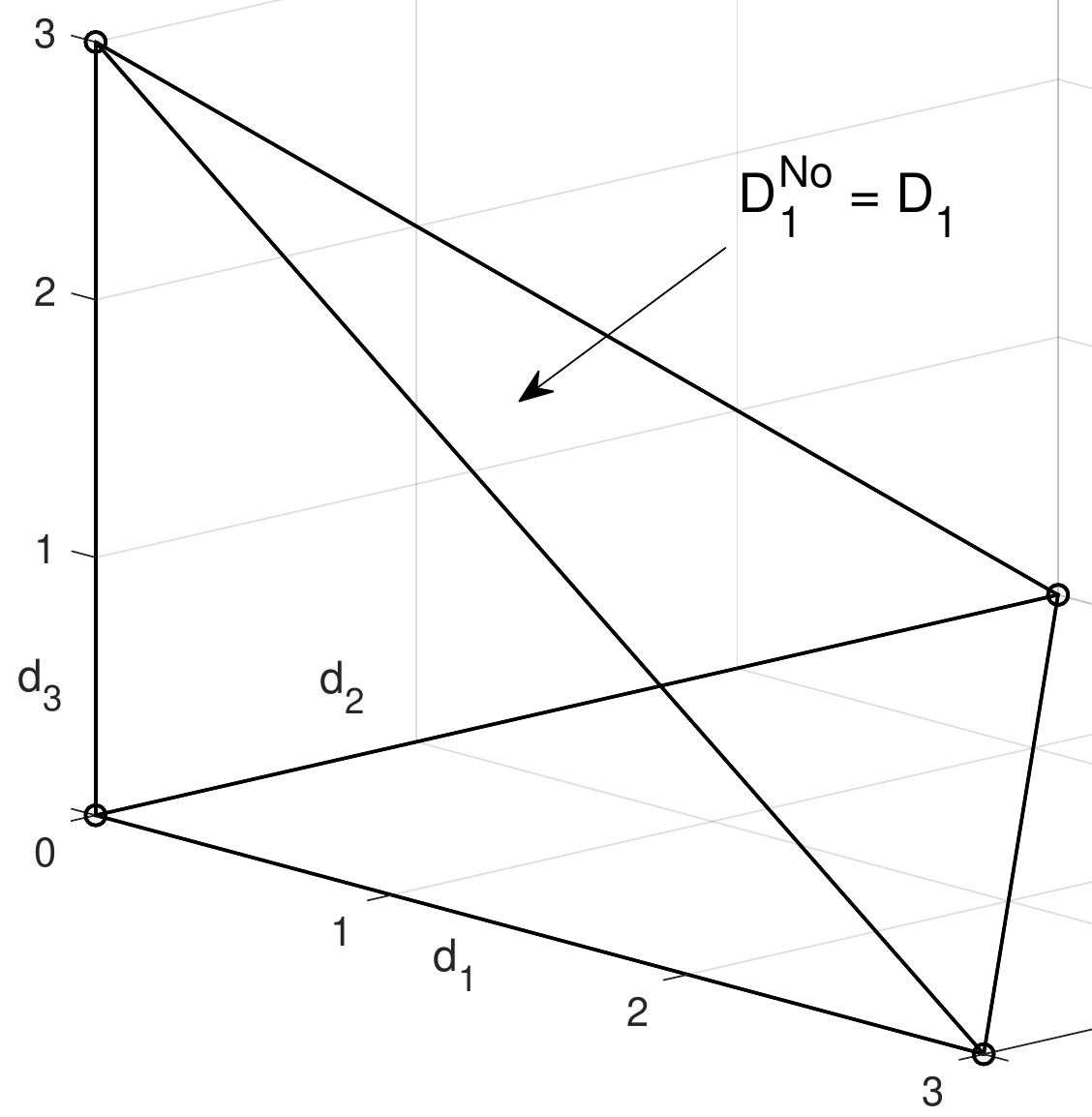}
		\caption{Order-1 messages: $(N_1,N_2,N_3,M) = (3,3,3,3).$}
		\label{fig:side:c}
	\end{minipage}%
	\begin{minipage}{0.5\linewidth}
		\centering
		\includegraphics[width=1.65in]{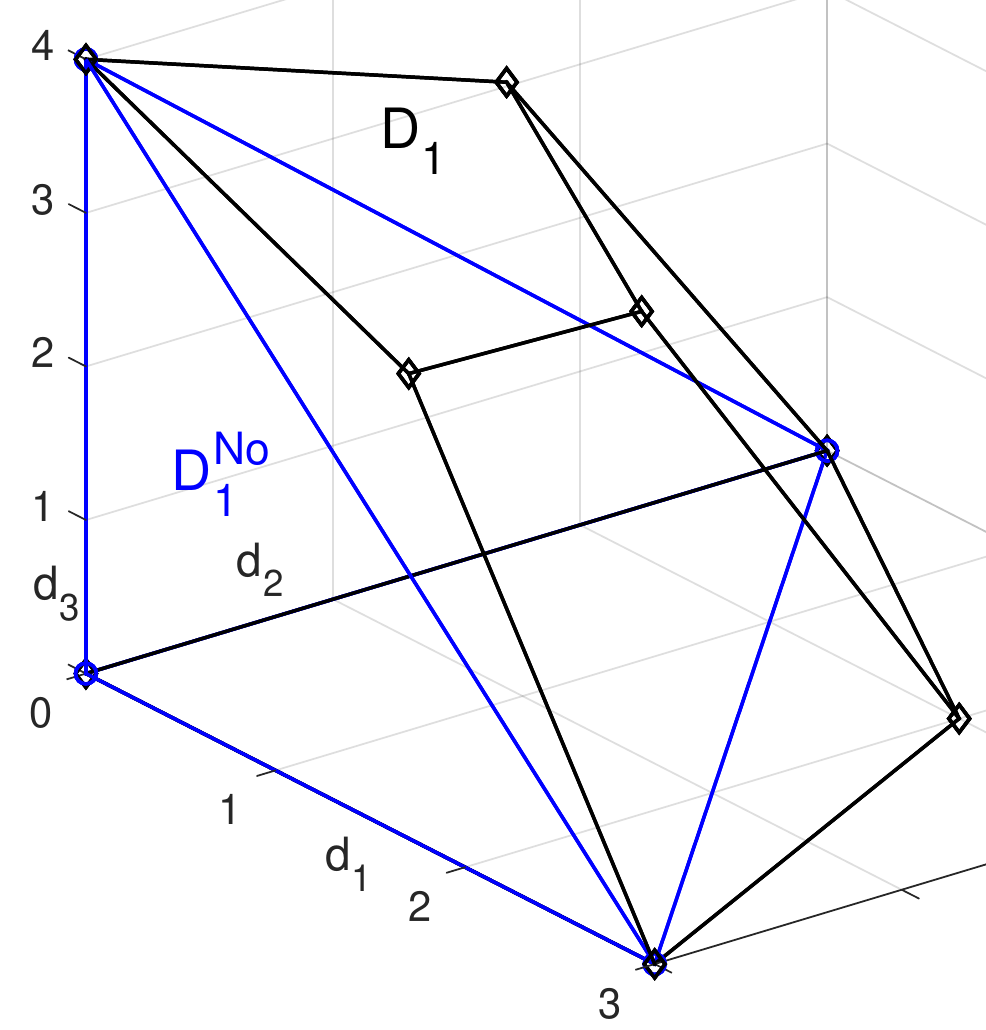}
		\caption{Order-1 messages: $(N_1,N_2,N_3,M) = (3,3,4,7).$}
		\label{fig:side:d}
	\end{minipage}
\end{figure}

For order-2 messages, the DoF region with no CSIT is derived by the corollary of Theorem 1 in \cite{53}, which is denoted by ${\cal{D}}_2^\text{No}$. Examined inequalities of ${\cal{D}}_2^\text{No}$ and ${\cal{D}}_2$\footnote{${\cal{D}}_2$ is given by Theorem 1 in this paper.}, if $M \le N_2$, it can be verified that  ${\cal{D}}_2^\text{No} = {\cal{D}}_2$. Otherwise, we have ${\cal{D}}_2^\text{No} \subset  {\cal{D}}_2$. In other words, for order-2 messages, the delayed CSIT is useful when $N_2 < M$. In Fig. 5, we provide an example for $M \le N_2$. In Fig. 6, we provide an example for $N_2<M$. 
	
For order-1 messages, according to \cite{53}, the DoF region with no CSIT is given by ${\cal{D}}_1^\text{No} = \{d_1,d_{2},d_3 \in \mathbb{R}_+^3| d_1/\min\{M,N_1\} + d_2/\min\{M,N_2\} + d_3/\min\{M,N_3\} \le 1\}$. We compare the results in \cite{9} and this paper with ${\cal{D}}_1^\text{No}$, by examining inequalities of ${\cal{D}}_1^\text{No}$ and ${\cal{D}}_1^\text{ach.}$. When $M \le N_2$, it can be verified that ${\cal{D}}_1^\text{No} = {\cal{D}}_1$\footnote{${\cal{D}}_1$ denotes the DoF region of three-user MIMO broadcast channel with delayed CSIT and order-1 messages.}. Otherwise,  we have ${\cal{D}}_1^\text{No} \subset {\cal{D}}_1^\text{ach.} \subseteq {\cal{D}}_1$. In other words, for order-1 messages, the delayed CSIT is useful when $N_2 < M$. In Fig. 7, we provide an example for $M \le N_2$. In Fig. 8, we provide an example for $N_2<M$.

\section{Conclusions}

For the three-user MIMO broadcast channel with delayed CSIT and arbitrary antenna configurations, we have characterized the DoF region of order-2 messages for arbitrary antenna configurations. We have also  obtained an achievable DoF of order-1 messages for $\max\{N_1 + N_2, N_3\} < M$ antenna configurations, which complements the insufficiency of the existing results. The results of this paper rely on the design of transmission scheme and the transformation approach. In particular, the transformation approach can systematically analyze the achievable DoF region of transmission scheme, whereas traditional methods cannot. Since the decoding condition of transmission scheme with delayed CSIT exists in general, the transformation approach has potential applications in the $K$-user MIMO broadcast channel and the other multi-user channels. In the future, two applications of our transformation approach will be elaborated, which are given as follows: 1) Appending the fresh order-3 message transmission in the last phase of our transmission schemes and applying the transformation approach, we can obtain the achievable DoF region of three-user MIMO broadcast channel with delayed CSIT, private and common messages; and 2) Generalizing our achievability and converse of order-2 messages to that of order-$(K-1)$ messages, we can derive the DoF region of $K$-user MIMO broadcast channel with delayed CSIT and order-$(K-1)$ messages.

\begin{appendices}

\section{Proof of Lemma 3}

The idea of proving this Lemma borrows from that of proving the Theorem 1 in \cite{53}. Similar to \cite{53}, a genie enhances the broadcast channel by providing $W_{12}$ and $W_{13}$ to receiver 2, and $W_{12},W_{23},W_{13}$ to receiver 3, where the order-2 message desired by receivers 1 and 2 is denoted by $W_{12}$, the order-2 message desired by receivers 2 and 3 is denoted by $W_{23}$, and the order-2 message desired by receivers 1 and 3 is denoted by $W_{13}$. According to Fano's inequality, the data rates of messages $W_{12},W_{13}$, and $W_{23}$ are bounded by
\begin{subequations}
	\begin{eqnarray}
		&&	n R_{12} + nR_{13} \le I(W_{12},W_{13};\textbf{y}_1^n|\textbf{H}^n ) + o(\log \text{SNR}), \label{Fano1} \\		
		&&	n R_{23}  \le I(W_{23};\textbf{y}_2^n|\textbf{H}^n, W_{12},W_{13}) + o(\log \text{SNR}), \label{Fano2}
	\end{eqnarray}
\end{subequations}
where the collection across $n$ channel uses of output signals at receiver $i,i=1,2$, and CSI matrices are denoted by $\textbf{y}_i^n$, and $\textbf{H}^n$, respectively; and $ \lim_{\text{SNR} \rightarrow {\cal{1}}} \frac{o(\log \text{SNR})}{\log \text{SNR}} = 0$. Based on \eqref{Fano1} and \eqref{Fano2}, we have
\begin{eqnarray}
	&& \frac{nR_{12}+ nR_{13}}{\min\{M,N_1\}} + \frac{nR_{23}}{\min\{M,N_2\}} \nonumber \\
	&& \overset{(a)}{\le} \frac{I(W_{12},W_{13};\textbf{y}_1^n|\textbf{H}^n)}{\min\{M,N_1\}} + \frac{I(W_{23};\textbf{y}_2^n|\textbf{H}^n, W_{12},W_{13})}{\min\{M,N_2\}}  + o(\log \text{SNR})  \nonumber \\
	&& =  \frac{h(\textbf{y}_1^n|\textbf{H}^n )}{\min\{M,N_1\}}	- \frac{h(\textbf{y}_1^n|\textbf{H}^n, W_{12},W_{13})}{\min\{M,N_1\}} + \frac{h(\textbf{y}_2^n|\textbf{H}^n, W_{12},W_{13})}{\min\{M,N_2\}}  - \frac{h(\textbf{y}_2^n|\textbf{H}^n, W_{12},W_{13},W_{23})}{\min\{M,N_2\}} \nonumber \\
	&& + o(\log \text{SNR}) \nonumber \\
	&& \overset{(b)}{=} \frac{h(\textbf{y}_1^n|\textbf{H}^n )}{\min\{M,N_1\}} + \frac{h(\textbf{y}_2^n|\textbf{H}^n, W_{12},W_{13})}{\min\{M,N_2\}}- \frac{h(\textbf{y}_1^n|\textbf{H}^n, W_{12},W_{13})}{\min\{M,N_1\}} + o(\log \text{SNR}) \nonumber \\
	&& \overset{(c)}{\le} \frac{h(\textbf{y}_1^n|\textbf{H}^n)}{\min\{M,N_1\}} +   o(\log \text{SNR}) \le n\log \text{SNR} + o(\log \text{SNR}), \label{noCSIT}
\end{eqnarray}
where (a) is from applying \eqref{Fano1}-\eqref{Fano2}; (b) is from $h(\textbf{y}_2^n|\textbf{H}^n, W_{12},W_{13},W_{23}) = o(\log \text{SNR})$; and (c) is from $\frac{h(\textbf{y}_2^n|\textbf{H}^n, W_{12},W_{13})}{\min\{M,N_2\}}- \frac{h(\textbf{y}_1^n|\textbf{H}^n, W_{12},W_{13})}{\min\{M,N_1\}}$ $\le 0$ by Lemma and (18) in \cite{53}. We can obtain a DoF outer region, having the same expression as \eqref{Col}, by dividing both sides of \eqref{noCSIT} by $n \log \text{SNR}$ and taking the limit of $n$. This DoF outer region can be achieved by the same way in the proof of Lemma 1. This completes the proof.

\section{General Transmission Scheme for Order-2 Symbols}

In the first step of Phase-I, the order-2 symbols $\textbf{x}_{ab} \in \mathbb{C}^{T_{12}B_1}$ desired by receivers 1 and 2 are transmitted with $T_{12}$ TSs and $B_1$ transmit antennas \footnote{The vector of order-2 symbols desired by receivers 1 and 2 is denoted by $\textbf{x}_{ab}$. The vector of order-2 symbols desired by receivers 2 and 3 is denoted by $\textbf{x}_{bc}$. The vector of order-2 symbols desired by receivers 1 and 3 is denoted by $\textbf{x}_{ac}$. The vector of order-3 symbols desired by receivers 1, 2, and 3 is denoted by $\textbf{x}_{abc}$.}. The received signals are given by
\begin{equation}
	\textbf{y}_i^{ab}= 
	\underbrace{\text{blkdiag}\{ 
		\textbf{H}_i[1], \cdots, \textbf{H}_i[T_{12}]
		\}}_{\textbf{H}_i^{ab}}  \textbf{x}_{ab}, \quad i =1,2,3.
\end{equation}
In the second step of Phase-I, the order-2 symbols $\textbf{x}_{bc} \in \mathbb{C}^{T_{23}B_2}$ desired by receivers 2 and 3 are transmitted with $T_{23}$ TSs and $B_2$ transmit antennas. The received signals are given by
\begin{equation}
	\textbf{y}_i^{bc}= 
	\underbrace{\text{blkdiag}\{  
		\textbf{H}_i[T_{12} + 1],\cdots, \textbf{H}_i[T_{12} + T_{23}]
		\}}_{\textbf{H}_i^{bc}} \textbf{x}_{bc}, \quad i=1,2,3.
\end{equation} 
In the final step of Phase-I, the order-2 symbols $\textbf{x}_{ac} \in \mathbb{C}^{T_{13}B_3}$ desired by receivers 1 and 3 are transmitted with $T_{13}$ TSs and $B_3$ transmit antennas. The received signals are given by
\begin{equation}
	\textbf{y}_i^{ac}= 
	\underbrace{\text{blkdiag}\{ 
		\textbf{H}_i[T_{12} + T_{23} + 1], \cdots, \textbf{H}_i[T_{12} + T_{23} + T_{13}]
		\}}_{\textbf{H}_i^{ac}}  \textbf{x}_{ac}, \quad i=1,2,3.
\end{equation}
If the order-2 symbols cannot be decoded  instantaneously, we design order-3 symbols to facilitate the decoding of transmitted order-2 symbols. Based on the Phase-I CSIT, the design of such order-3 symbols can be given by
\begin{equation}  \label{O3}
	\textbf{x}_{abc} 
	=  \begin{bmatrix} 
		\underline{\textbf{y}}_3^{ab} + \underline{\textbf{y}}_1^{bc}\\ 
		\underline{\textbf{y}}_2^{ac} + \underline{\textbf{y}}_1^{bc} 
	\end{bmatrix} \in \mathbb{C}^{TC},
\end{equation}
where $ \underline{\textbf{y}}_1^{bc} \in \mathbb{C}^{T_{23}(B_2 - N_2)},  \underline{\textbf{y}}_2^{ac} \in \mathbb{C}^{T_{13}(B_3 - N_1)}$, and $
\underline{\textbf{y}}_3^{ab}\in \mathbb{C}^{T_{12}(B_1 - N_1)}
$ are truncated vectors from $\textbf{y}_1^{bc} \in \mathbb{C}^{T_{23}N_1}, \textbf{y}_2^{ac} \in \mathbb{C}^{T_{13}N_2}$, and $\textbf{y}_3^{ab} \in \mathbb{C}^{T_{12}N_3}$, respectively. After decoding  $\textbf{x}_{abc}$, receiver 1 can acquire $\underline{\textbf{y}}_3^{ab}$ by cancellation $\underline{\textbf{y}}_3^{ab} + \underline{\textbf{y}}_1^{bc} - \underline{\textbf{y}}_1^{bc}$ and $\underline{\textbf{y}}_2^{ac}$
by cancellation $\underline{\textbf{y}}_2^{ac} + \underline{\textbf{y}}_1^{bc} - \underline{\textbf{y}}_1^{bc}$. After decoding $\textbf{x}_{abc}$, receiver 2 can acquire $\underline{\textbf{y}}_1^{bc}$ by cancellation $\underline{\textbf{y}}_2^{ac} + \underline{\textbf{y}}_1^{bc} - \underline{\textbf{y}}_2^{ac}$ and $\underline{\textbf{y}}_3^{ab}$ by cancellation $\underline{\textbf{y}}_3^{ab} + \underline{\textbf{y}}_1^{bc} - \underline{\textbf{y}}_1^{bc}$. After decoding $\textbf{x}_{abc}$, receiver 3 can acquire $\underline{\textbf{y}}_1^{bc}$ by cancellation $\underline{\textbf{y}}_3^{ab} + \underline{\textbf{y}}_1^{bc} - \underline{\textbf{y}}_3^{ab}$ and $\underline{\textbf{y}}_2^{ac}$ by cancellation $\underline{\textbf{y}}_2^{ac} + \underline{\textbf{y}}_1^{bc} - \underline{\textbf{y}}_1^{bc}$. Therefore, $\textbf{x}_{abc}$ are used for the decoding of order-2 symbols at all receivers, which are order-3 symbols.  In Phase-II,  order-3 symbols $\textbf{x}_{abc}$ are transmitted with $T$ TSs and $C$ antennas. To ensure the instantaneous decoding of order-3 symbols, we set $C = N_2$ if $N_2<M \le N_3$, and $C = N_3$ otherwise.

\section{Coded Transmission of Order-1 Symbols}

At the first step, $\textbf{x}_a^1 + \textbf{x}_b^1 \in {\mathbb{C}^{T_1A_1}}$ is transmitted using $T_{1}$ TSs and $A_1$ transmit antennas, which is a sum of receivers 1 and 2 desired symbols \footnote{The $i^\text{th}, i=1,2$ vector of order-1 symbols desired by receivers 1, 2, and 3 are denoted by $\textbf{x}^i_a$, $\textbf{x}^i_b$, $\textbf{x}^i_c$, respectively.}. The received signals are given by
 \begin{equation}
\textbf{y}_i^{a + b}= 
\underbrace{\text{blkdiag}\{ 
\textbf{H}_i[1], \cdots, \textbf{H}_i[T_1]
\}}_{\textbf{H}_i^{a + b}}  \left(\textbf{x}_a^1 + \textbf{x}_b^1\right), \quad i=1,2,3.
\end{equation}
At the second step, $\textbf{x}_b^2 + \textbf{x}_c^1 \in {\mathbb{C}^{T_2A_2}}$ is transmitted using $T_2$ TSs and $A_2$ transmit antennas, which is a sum of receivers 2 and 3 desired symbols. The received signals are given by
\begin{equation}
\textbf{y}_i^{b + c}= 
\underbrace{\text{blkdiag}\{  
\textbf{H}_i[T_1 + 1], \cdots,  \textbf{H}_i[T_1 + T_2]
\}}_{\textbf{H}_i^{b + c}}  \left(\textbf{x}_b^2 + \textbf{x}_c^1\right), \quad i=1,2,3.
\end{equation}
At the final step, $\textbf{x}_a^2 + \textbf{x}_c^2 \in {\mathbb{C}^{T_3A_3}}$ is transmitted using $T_3$ TSs and $A_3$ transmit antennas, which is a sum of receivers 1 and 3 desired symbols. The received signals are given by
    \begin{equation}
\textbf{y}_i^{a + c}= 
\underbrace{\text{blkdiag}\{  
\textbf{H}_i[T_1 + T_2 + 1], \cdots, \textbf{H}_i[T_1 + T_2 + T_3]\}}_{\textbf{H}_i^{a + c}}  \left(\textbf{x}_a^2 + \textbf{x}_c^2\right), \quad i=1,2,3.
\end{equation}

To provide equations in order to decode the transmitted order-1 symbols, based on the CSIT from the $1^{st}$ to the $(T_1 + T_2 + T_3)^{th}$ TS, we design the following order-2 symbols:
\begin{itemize}
\item This design aims to provide $\min\{A_1,N_1 + N_2\}T_1$ equations about $\textbf{x}_a^1$  to receiver 1  and $\min\{A_1,N_1 + N_2\}T_1$ equations about $\textbf{x}_b^1$  to receiver 2. Given $\underline{\textbf{H}}_2^{a + b}\textbf{x}_a^1 \in \mathbb{C}^{(\min\{A_1,N_1 + N_2\}-N_1)T_1}$ to receivers 1 and 2, then receiver 1 will acquire $(\min\{A_1,N_1 + N_2\}-N_1)T_1$ equations, and receiver 2 will acquire $(\min\{A_1,N_1 + N_2\}-N_1)T_1$ equations as well by $\underline{\textbf{y}}_2^{a + b} - \underline{\textbf{H}}_2^{a + b}\textbf{x}_a^1 \in \mathbb{C}^{(\min\{A_1,N_1 + N_2\}-N_1)T_1}$, due to $\min\{A_1,N_1 + N_2\} - N_1 \le N_2$. If we provide $\textbf{H}_1^{a + b}\textbf{x}_b^1 \in \mathbb{C}^{N_1T_1}$ to receivers 1 and 2, then receiver 1 will obtain $N_1T_1$ new equations by $\textbf{y}_1^{a + b} - \textbf{H}_1^{a + b}\textbf{x}_b^1$, and receiver 2 will obtain $N_1T_1$ new equations as well. The generated order-2 symbols are 
\begin{equation} 
\textbf{x}_{ab} 
=  \begin{bmatrix}
\underline{\textbf{H}}_2^{a + b}\textbf{x}_a^1 \\
\textbf{H}_1^{a + b}\textbf{x}_b^1
\end{bmatrix} \in \mathbb{C}^{\min\{A_1,N_1 + N_2\}T_1}.
\end{equation}
\item  This design aims to provide $\min\{A_2,N_2 + N_3\}T_2$ equations about $\textbf{x}_b^2$ to receiver 2  and $\min\{A_2,N_2 + N_3\}T_2$ equations about $\textbf{x}_c^1$  to receiver 3. Given $\underline{\textbf{H}}_3^{b + c}\textbf{x}_b^2 \in \mathbb{C}^{(\min\{A_2,N_2 + N_3\}-N_2)T_2}$ to receivers 2 and 3, then receiver 2 will acquire $(\min\{A_2,N_2 + N_3\}-N_2)T_2$ equations, and receiver 3 will acquire $(\min\{A_2,N_2 + N_3\}-N_2)T_2$ equations as well by $\underline{\textbf{y}}_3^{b + c} - \underline{\textbf{H}}_3^{b + c}\textbf{x}_b^2 \in \mathbb{C}^{(\min\{A_2,N_2 + N_3\}-N_2)T_2}$, due to $\min\{A_2,N_2 + N_3\} -N_2 \le N_3$. If we provide $\textbf{H}_2^{b + c}\textbf{x}_c^1 \in \mathbb{C}^{N_2T_2}$ to receivers 2 and 3, then receiver 2 will obtain $N_2T_2$ new equations by $\textbf{y}_2^{b + c} - \textbf{H}_2^{b + c}\textbf{x}_c^1$, and receiver 3 will obtain $N_2T_2$ new equations as well.  The generated order-2 symbols are 
\begin{equation}
\textbf{x}_{bc} 
=  \begin{bmatrix}
\underline{\textbf{H}}_3^{b + c}\textbf{x}_b^2 \\
\textbf{H}_2^{b + c}\textbf{x}_c^1
\end{bmatrix} \in \mathbb{C}^{\min\{A_2,N_2 + N_3\}T_2}.
\end{equation}
\item  This design aims to provide $\min\{A_3,N_1 + N_3\}T_3$ equations about $\textbf{x}_a^2$ to receiver 1  and $\min\{A_3,N_1 + N_3\}T_3$ equations about $\textbf{x}_c^2$  to receiver 3. Given $\underline{\textbf{H}}_3^{a + c}\textbf{x}_a^2 \in \mathbb{C}^{(\min\{A_3,N_1 + N_3\}-N_1)T_3}$ to receivers 1 and 3, then receiver 1 will acquire $(\min\{A_3,N_1 + N_3\}-N_1)T_3$ equations, and receiver 3 will acquire $(\min\{A_3,N_1 + N_3\}-N_1)T_3$ equations as well by $\underline{\textbf{y}}_3^{a + c} - \underline{\textbf{H}}_3^{a + c}\textbf{x}_a^2 \in \mathbb{C}^{(\min\{A_3,N_1 + N_3\}-N_1)T_3}$, due to $\min\{A_3,N_1 + N_3\} - N_1 \le N_3$. If we provide $\textbf{H}_1^{a + c}\textbf{x}_c^2 \in \mathbb{C}^{N_1T_3}$ to receivers 1 and 3, then receiver 1 will obtain $N_1T_3$ new equations by $\textbf{y}_1^{a + c} - \textbf{H}_1^{a + c}\textbf{x}_c^2$, and receiver 3 will obtain $N_1T_3$ new equations as well.  The generated order-2 symbols are 
\begin{equation} 
\textbf{x}_{ac} 
=  
\begin{bmatrix}
\underline{\textbf{H}}_3^{a + c}\textbf{x}_a^2 \\
\textbf{H}_1^{a + c}\textbf{x}_c^2
\end{bmatrix} \in \mathbb{C}^{\min\{A_3,N_1 + N_3\}T_3}.
\end{equation}
\end{itemize}

\end{appendices}

\bibliographystyle{IEEEtran}
\bibliography{DoF} 

\end{document}